\documentclass[%
superscriptaddress,
showpacs,preprintnumbers,
nofootinbib,
 amsmath,amssymb,
 aps,
 twocolumn,
 prd,
]{revtex4-1}
\usepackage{braket}
\usepackage{bm}
\usepackage{dcolumn}
\usepackage{cancel}
\usepackage[dvipdfmx,final]{graphicx}
\input{colordvi.tex}
\usepackage{longtable}
\usepackage[compat=1.1.0]{tikz-feynman}
\usepackage{mathtools}
\usepackage{tabularx}
\tikzfeynmanset{every edge={very thick},}\usepackage[pdftex,bookmarks,linktocpage,pdfpagelabels,plainpages=false,hyperfigures,linkcolor=blue,citecolor=blue]{hyperref}
\usepackage{cleveref}
\hypersetup{colorlinks=true}
\usepackage{color}

\begin{document}
\preprint{CTPU-PTC-22-19}
\title{Evolution of Resonant Self-interacting Dark Matter Halos}
\author{Ayuki Kamada}
\email{akamada@fuw.edu.pl}
\affiliation{Institute of Theoretical Physics, Faculty of Physics, University of Warsaw, ul. Pasteura 5, PL–02–093 Warsaw, Poland}
\author{Hee Jung Kim}
\email{heejungkim@ibs.re.kr}
\affiliation{Center for Theoretical Physics of the Universe, Institute for Basic Science (IBS), Daejeon 34126, Korea}
\date{\today}

\begin{abstract}
Recent analysis on the stellar kinematics of ultra-faint dwarf (UFD) galaxies has put a stringent upper limit on the self-scattering cross section of dark matter, i.e., $\sigma/m<{\cal O}(0.1)\,{\rm cm^2/g}$ at the scattering velocity of ${\cal O}(10)\,{\rm km/s}$.
Resonant self-interacting dark matter (rSIDM) is one possibility that can be consistent with the UFDs and explain the low central densities of rotation-supported galaxies;
the cross section is resonantly enhanced to be $\sigma/m = {\cal O}(1)\,{\rm cm^2/g}$ around the scattering velocity of ${\cal O}(100)\,{\rm km/s}$ while being suppressed at lower velocities.
To further assess this possibility, since the inferred dark matter distribution of halos from astrophysical observations is usually compared to that in constant-cross section SIDM (cSIDM), whether the structures of rSIDM halos can be approximated by the cSIDM halo profiles needs to be clarified.
In this work, we employ the grovothermal fluid method to investigate the structural evolution of rSIDM halos in a wide mass range.
We find that except for halos in a specific mass range, the present structures of rSIDM halos are virtually indistinguishable from those of the cSIDM halos.
For halos in the specific mass range, the resonant self-scattering renders a break in their density profile.
We demonstrate how such a density-profile break appears in astrophysical observations, e.g., rotation curves and line-of-sight velocity dispersion profiles.
We show that for halos above the specific mass range, the density-profile break thermalizes to disappear before the present.
We demonstrate that such distinctive thermalization dynamics can leave imprints on the orbital classes of stars with similar ages and metallicities.
\end{abstract}

\maketitle

\section{Introduction}

Collisionless cold dark matter (CDM) is a successful paradigm in explaining the observed structures at large scales.
At small scales ($\lesssim 1\,{\rm Mpc}$), however, there are reported discrepancies between predictions of CDM-only simulations and astrophysical observations (see, e.g., \cite{Bullock:2017xww} for a review on the small-scale issues).
For example, the observed rotation curves in dark-matter (DM) dominated galaxies seem to prefer a flat central density distribution~\cite{deBlok:2001hbg,Salucci:2007tm,Oh:2010ea,Oh:2015xoa}, i.e., a core, rather than the cuspy profile predicted in collisionless CDM-only simulations~\cite{Navarro:1995iw,Navarro:1996gj}.
The sub-grid astrophysical processes of baryons, e.g., supernova feedback and stellar winds, are known to play a role in resolving the issues~\cite{Navarro:1996bv,Governato:2009bg,Pontzen:2011ty,Teyssier:2012ie,Pontzen:2014lma,Sawala:2015cdf,Dutton:2015nvy,Wetzel:2016wro,Read:2015sta}, while it is not clear to what degree the baryonic physics affects the halo evolution~\cite{Pontzen:2011ty,Benitez-Llambay:2019wfi}.

In light of this situation, the possibility of DM microphysics being responsible for the small-scale issues has gained considerable attention, resonating with new and creative experimental efforts to test new DM benchmarks beyond the traditional weakly interacting massive particles~\cite{Graham:2015ouw,Battaglieri:2017aum}.
Dedicated studies are demonstrating how DM microphysics can be probed by looking into the spatial distribution of DM~\cite{Buckley:2017ijx,Marsh:2021lqg}.
Such gravitational probes of DM are appealing since we may probe DM interactions secluded from the Standard Model (SM), which are difficult to probe in terrestrial experiments.
In order to maximally utilize the upcoming and archival data, it is important to sharpen the predictions of DM microphysics on the spatial distribution of DM and identify the distinctive features that only a specific DM microphysics would exhibit.

Self-interacting dark matter (SIDM) is a promising framework that mitigates some of the most prevalent aspects of the small-scale issues (see \cite{Tulin:2017ara,Adhikari:2022sbh} for reviews).
The elastic self-scattering with cross section per DM mass of $\sigma/m\sim 1\, {\rm cm^2/g}$ induces macroscopic conduction of DM kinetic energy, which leads to thermalization of DM particles in the central region of a halo in the timescale of the age of the Universe.
Such thermalization results in the formation of a low-density core in DM-dominated galaxies~\cite{Zavala:2012us,Elbert:2014bma}, or a denser and smaller core in baryon-rich galaxies~\cite{Kamada:2016euw}, respective aspect alleviating the core-cusp~\cite{Moore:1999nt,deBlok:2009sp} (and too-big-to-fail~\cite{Boylan-Kolchin:2011qkt,Boylan-Kolchin:2011lmk}) issue, and the diversity issue~\cite{Oman:2015xda}.
The former aspect is most highlighted for dwarf/low surface brightness (LSB) spiral galaxies ($10^{9}$-$10^{12}\,{\rm M_\odot}$), showing the preference of $\sigma/m \simeq 2\,{\rm cm^2/g}$ at the DM scattering velocities of $30$-$200\,{\rm km/s}$~\cite{Kaplinghat:2015aga}.
On the other hand, a smaller cross section seems to be preferred at larger velocities:
observations on galaxy clusters ($\sim10^{14}\,{\rm M_\odot}$) constrain the cross section to be $\sigma/m\lesssim 0.1\,{\rm cm^2/g}$ for velocities larger than $\gtrsim1000\,{\rm km/s}$~\cite{Kaplinghat:2015aga,Randall:2008ppe,Harvey:2018uwf,Sagunski:2020spe,Andrade:2020lqq}.

Milky Way's ultra-faint dwarf (UFD) galaxies are one of the ideal sites for probing the SIDM cross section in the low-velocity regime ($\lesssim 30\,{\rm km/s}$), since their central DM density profile is expected to be less affected by astrophysical processes~\cite{Tollet:2015gqa,Fitts:2016usl,Lazar:2020pjs} and more accurate information about their structure will be provided by next-generation spectroscopic surveys in near future~\cite{EUCLID:2011zbd,PFSTeam:2012fqu}.
Recently, the density profiles of UFDs have been analyzed assuming the core expansion (formation) phase of SIDM~\cite{Hayashi:2020syu}.
Among the 23 considered UFDs, two of them, i.e., Segue 1 and Willman 1, put a stringent upper bound on $\sigma/m\lesssim 0.1\,{\rm cm^2/g}$ in the low-velocity regime.
Further investigation may be warranted since the limited stellar kinematic data within small radii of the UFDs may render a large uncertainty in their structural parameters.
Nevertheless, their result points to an interesting velocity dependence of $\sigma/m$ when we take the constraint at face values.
Along with the constraints on $\sigma/m$ from galaxy clusters, $\sigma/m$ is constrained to be $\lesssim 0.1\,{\rm cm^2/g}$ at the velocity scales of $\gtrsim 1000\,{\rm km/s}$ and $\lesssim 30\,{\rm km/s}$, while field dwarf/LSB galaxies prefer a larger value $\sim2\,{\rm cm^2/g}$ at the intermediate velocity scales~(see also Fig.~\ref{fig:rSIDM}).
This requires an enhancement of $\sigma/m$ around $\sim 100\,{\rm km/s}$ and a sharp drop by an order of magnitude towards lower velocities around $\sim 30\,{\rm km/s}$.
It is interesting to ask which particle physics realization exhibits such velocity dependence.

The required velocity dependence can be realized when DM particles self-scatter through a resonant intermediate state~\cite{Chu:2018fzy}.
When the mass of the resonance is just above twice of the DM mass, the self-scattering is resonantly enhanced around some scattering velocity $v_R$.
Such near-threshold resonances exist in SM QCD, and QCD-like theories of DM can realize such states~\cite{Tsai:2020vpi,Kondo:2022lgg}.
For systems of velocities away from the $v_R$, e.g., galaxy clusters and UFDs, the self-scattering misses the resonance and hence exhibits smaller values of $\sigma/m$.
Moreover, the transition width of the velocity-averaged cross section towards lower velocity can be sharp enough to be consistent with the constraints from the UFDs, as will be discussed in the next section.

Although resonant SIDM (rSIDM) appears to accommodate the velocity dependence consistent with the observations of UFDs and field dwarf/LSBs, there has been no explicit study on the evolution of rSIDM halos.
Most importantly, it has not been clarified if structures of rSIDM halos can be mapped onto that in scenarios of constant SIDM (cSIDM) cross section;
since the favored values and upper bounds on the SIDM cross section at a given scattering velocity are usually inferred by applying the isothermal Jeans modeling for cSIDM halos~\cite{Kaplinghat:2013xca,Kaplinghat:2015aga,Jiang:2022aqw} to observations (including the constraints from the UFDs~\cite{Hayashi:2020syu}), it is necessary to check to what extent the existing results for cSIDM apply to rSIDM.
For the velocity dependence of Coulomb/Yukawa-like self-interactions, the application of the isothermal Jeans modeling to simulated velocity-dependent SIDM halos has been explored~\cite{Robertson:2020pxj}, and a method for the mapping has been proposed by employing the gravothermal fluid equations~\cite{Yang:2022hkm}, making it possible for the velocity-dependent SIDM model parameters to be constrained from astrophysical observations.
The mapping of halo structures may not exist for rSIDM halos at the vicinity of the resonant velocity;
inside an rSIDM halo, only a specific region may exhibit the enhanced heat conduction due to the sharp velocity dependence of $\sigma/m$.
Such selective enhancement inside an rSIDM halo could result in a distinctive halo structure at present.

The goal of this study is to provide the first step in understanding the structural evolution of rSIDM halos.
By employing the gravothermal fluid method, we numerically follow the evolution of isolated halos in the core expansion phase.
This method allows one to follow the evolution of halo structures down to small radii, e.g., $r \lesssim 10\,{\rm pc}$ for UFDs, at a small computational cost, allowing us to accurately follow the distinctive evolution of rSIDM halo structures and scope the dependence of the evolution on rSIDM parameters and halo mass.
We take the benchmarks for $p$-wave resonant self-scattering that represents the possible velocity dependencies that may be consistent with the observations on the galaxies of different velocity scales~(see also the black curves in Fig.~\ref{fig:rSIDM}).~\footnote{Similar velocity dependencies are also possible in $s$-wave resonant scattering; see the discussion below Eq.~\eqref{eq:NWA}. We take the $p$-wave benchmarks for purely illustrative reasons as one of the $p$-wave benchmarks, i.e., ${\rm P}1$, was presented in Ref.~\cite{Chu:2018fzy}.}

Notably, for halos with scattering velocities close to the resonant velocity, a break in the density profile develops at the radius where the local heat conduction rate from DM self-scattering is resonantly enhanced.
Such a break in the density profile is eventually thermalized and the global density profile converges to that in cSIDM.
We find that for the $p$-wave rSIDM benchmarks, rSIDM halos in a specific mass range would exhibit the density break at present.
Such a density break can be explicitly seen in astrophysical observations, e.g., in stellar line-of-sight velocity dispersion (LOSVD) profiles of MW satellites and rotation curves of dwarf/LSB galaxies, and serve as a smoking-gun signature for rSIDM.
We also remark that the thermalization dynamics could exhibit a period during which the density break develops into a shock-like form that propagates toward the center.
For halos of mass larger than the aforementioned range, the propagation is complete at present and thus the distinction between cSIDM and rSIDM is not manifest in the density profiles.
Nevertheless, the propagation dynamics of the density break in rSIDM halos can still be observationally important since it may leave imprints on the orbital classes of stars formed around the time of the propagation.

In Section~\ref{section:gravofluid}, we begin with the parametrization of the self-scattering cross section in rSIDM and present the gravothermal fluid method for treating the velocity dependence of $\sigma/m$.
Our numerical results, showing the halo evolution for different benchmarks and halo masses, are presented in Section~\ref{section:rSIDMevolution}.
We also identify the halo-mass range where the density break can be seen at present.
In Section~\ref{section:observations}, we discuss the possible imprints of rSIDM halo dynamics on astrophysical observations.
We give concluding remarks in Section~\ref{section:conclusion}.

\section{Methodology} \label{section:gravofluid}

{\bf Resonant self-interaction.}
In the presence of a particle resonance mediating the self-scattering of DM, the (spin-averaged) non-relativistic cross section $\sigma$ can be parametrized as a sum of a constant piece $\sigma_0$ and resonant piece parametrized by a Breit-Wigner form~\cite{Chu:2018fzy}:
\begin{equation}
\sigma = \sigma_0 + \frac{4\pi S}{mE(v_{{\rm rel}})}\frac{\Gamma(v_{{\rm rel}})^{2}/4}{\left[E(v_{{\rm rel}})-E(v_{R})\right]^{2}+\Gamma(v_{{\rm rel}})^{2}/4}\,,
\label{eq:BWparam}
\end{equation} 
where $m$ is the DM mass, $E(v)=(m/2)v^2/2$, and $S=(2s_R+1)/(2s_{\rm dm}+1)^2$ is the symmetry factor taking into account the spin degrees of freedom of DM ($s_{\rm dm}$) and the resonance ($s_R$).
The resonant velocity is given as $E(v_R)=m_R-2m$ where $m_R$ is the mass of the resonance.
We assume that the total decay width of the resonance is dominated by $R\rightarrow {\rm dm}\, {\rm dm}$ around the resonant velocity;
we parametrize the momentum-dependent decay width as $\Gamma(v_{\rm rel})=m_R \gamma v_{\rm rel}^{2L+1}$, where $L$ is the orbital angular momentum for the self-scattering and $\gamma$ parametrizes the coupling between the resonance and DM.

\begin{figure}[t!]
\includegraphics[width=0.45\textwidth]{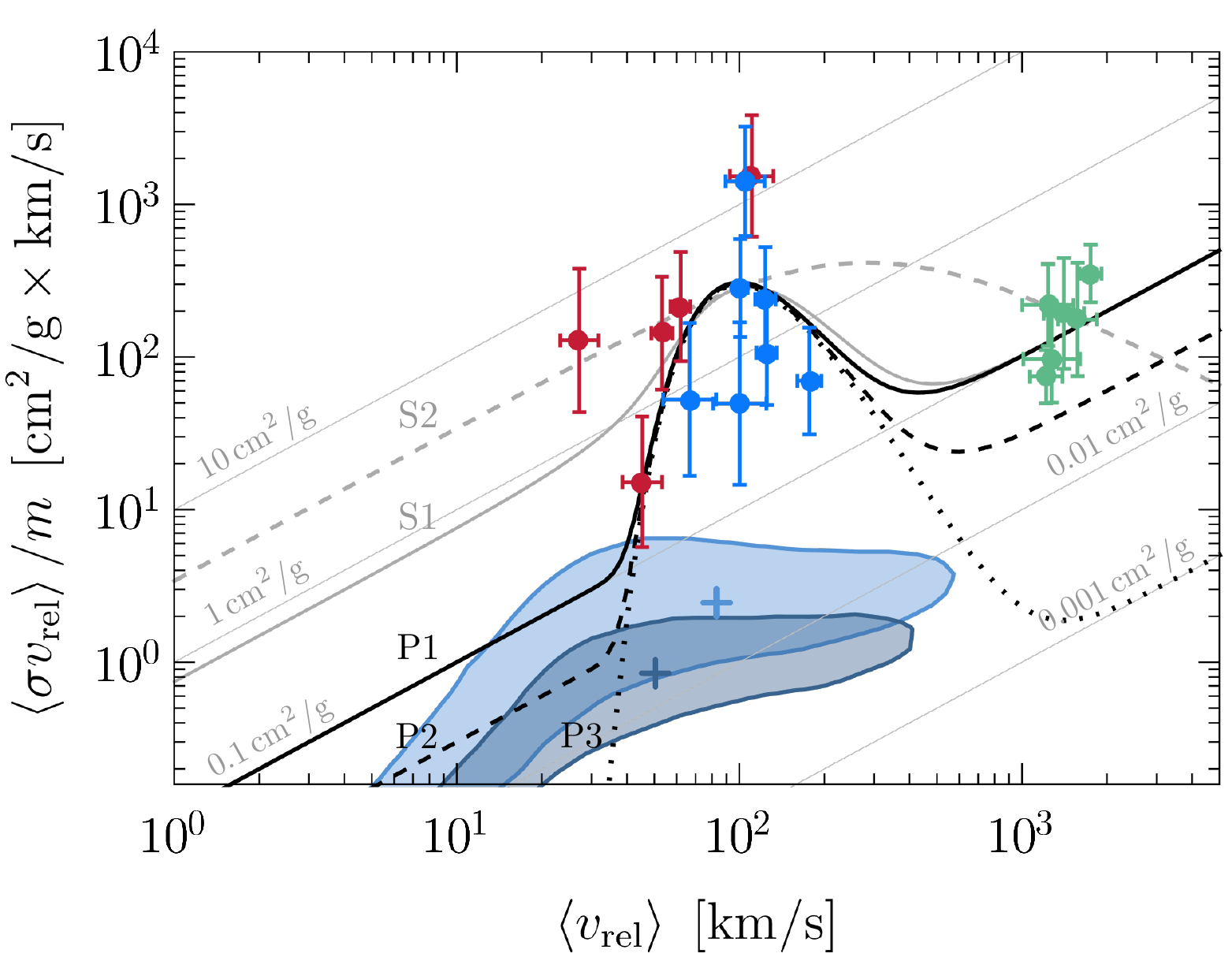} 
\caption{Velocity-weighted average of the resonant SIDM cross section per DM mass. The data points with error bars are the inferred SIDM cross sections from field dwarf (red)/LSB (blue) galaxies and galaxy clusters (green)~\cite{Kaplinghat:2015aga}; the curves labeled by ${\rm S1}$ (${\rm S2}$) and ${\rm P1}$ are the best-fit curves to the data points in the narrow (broad)-width $s$-wave and $p$-wave resonant scattering, respectively~\cite{Chu:2018fzy}. The colored regions are inferred (1$\sigma$) from UFDs (Willman 1 and Segue 1)~\cite{Hayashi:2020syu}.
Motivated from the stringent upper limit from the UFDs, we explore the ${\rm P2}$ and ${\rm P3}$ benchmark parameters; they are the same with the ${\rm P1}$ benchmark, but with smaller offset cross sections, i.e., $0.03\,{\rm cm^2/g}$ and $0.001\,{\rm cm^2/g}$, respectively.
}
\label{fig:rSIDM}
\end{figure}

Inside a halo, we approximate that the scattering velocity $v_{\rm rel}$ follows the Maxwell-Boltzmann distribution parametrized by the local one-dimensional velocity dispersion $\nu(r)$:
\begin{equation}
f(v_{\rm rel};\nu)=\frac{ v_{\rm rel}^2 }{ \sqrt{4\pi} \nu^3 }\exp\left( -\frac{ v_{\rm rel}^2 }{ 4\nu^2 } \right)\,.
\label{eq:MBdist}
\end{equation}
We will denote the distribution averaging by $\langle \cdot \rangle$; the integration range for $v_{\rm rel}$ is taken to be from $0$ to the local escape velocity which is usually larger than the local velocity dispersion in the central region of a halo.
In this work, we take the local escape velocity to be infinity.
Note that the expectation value of the scattering velocity is given as $\langle v_{\rm rel} \rangle = (4/\sqrt{\pi}) \nu$.
The semi-analytic method of isothermal Jeans modeling is often used to fit the predicted cSIDM halo profile to the observed astrophysical data in the core expansion phase~\cite{Kaplinghat:2013xca,Kaplinghat:2015aga,Kamada:2016euw,Valli:2017ktb,Robertson:2020pxj};
there, the quantity inferred from observations is $\langle \sigma v_{\rm rel} \rangle/m$ (vertical axis of Fig.~\ref{fig:rSIDM}) at a given DM scattering velocity (one-dimensional velocity dispersion) that characterizes the isothermal profile for the inner core (horizontal axis of Fig.~\ref{fig:rSIDM}).
As shown in Fig.~\ref{fig:rSIDM}, the dwarf/LSB galaxies (red/blue data points) prefer $\langle \sigma v_{\rm rel} \rangle / (m \langle v_{\rm rel} \rangle) \sim 2\,{\rm cm^2/g}$ around $\langle v_{\rm rel} \rangle \sim 100\,{\rm km/s}$, while the UFDs put a stringent upper bound as $\lesssim 0.1\,{\rm cm^2/g}$ at low velocities, i.e., $\lesssim 30\,{\rm km/s}$.

Such a sharp drop towards lower velocities is realized in rSIDM in the limit of a narrow resonance width.
The resonant contribution, i.e., the second term in the RHS of Eq.~\eqref{eq:BWparam}, to $\langle \sigma v_{\rm rel} \rangle/m$ around $v_{\rm rel}=v_R$ can be picked up by using a narrow-width approximation (NWA)~\cite{Chu:2018fzy}:
\begin{equation}
\frac{\langle\sigma v_{\rm rel}\rangle}{m}\bigg|_{\rm res.}=  \frac{16\pi^{3/2} S \gamma v_R^{2L+1}}{m^3\nu^3}e^{-\frac{v_R^2}{4\nu^2}}\,,
\label{eq:NWA}
\end{equation}
which is a good estimation of $\langle \sigma v_{\rm rel} \rangle/m$ around the resonance for $\gamma v_R^{2L-1}\lesssim 1$.
The resonant part in the narrow-width limit exhibits the minimal transition width towards lower velocities, $\Delta \langle v_{\rm rel}\rangle \sim v_R/2$;
hereafter, we will focus on this case.
The peak of the distribution-averaged cross section given in Eq.~\eqref{eq:NWA} happens at the $\langle v_{\rm rel} \rangle = \sqrt{8/3\pi} \, v_R$.

In Fig.~\ref{fig:rSIDM}, we display the velocity dependence of $\langle \sigma v_{\rm rel} \rangle/m$ for the benchmark parameters that fit the observations on dwarf/LSB galaxies and galaxy clusters, i.e., ${\rm S1}$, ${\rm S2}$ and ${\rm P1}$~\cite{Chu:2018fzy}.
The ${\rm S1}$ (${\rm S2}$) benchmark represents the case of narrow (broad) $s$-wave resonance, i.e., $L=0$;
the rSIDM parameters for the ${\rm S1}$ (${\rm S2}$) benchmark are $v_R=120\,{\rm km/s}$ ($5035\,{\rm km/s}$), $\gamma=10^{-4.5}$ ($10^{-1.1}$), $m/S^{1/3}=22\,{\rm GeV}$ ($16\,{\rm GeV}$) and $\sigma_0/m=0.1\,{\rm cm^2/g}$ ($\ll 0.1\,{\rm cm^2/g}$).
Away from the resonant velocities, the non-vanishing Breit-Wigner distribution renders out-of-pole contributions which have additional $\gamma$-suppression compared to the resonant one~(see Appendix~\ref{appendix:averages} for more discussion).
The low-velocity limit of the out-of-pole contribution to $\langle \sigma v_{\rm rel} \rangle/(m\langle v_{\rm rel} \rangle)$ is $\sim 2^{4(L+1)} \pi \gamma^2 \nu^{4L} / (m^3 v_R^4)$ which is not velocity-suppressed for the $s$-wave scattering.
Such out-of-pole values can be larger than the taken offset value $\sigma_0 \langle v_{\rm rel} \rangle/m$.
Nevertheless, one can always find $\gamma$ and $m$ where the out-of-pole values ($\propto\gamma^2/m^3$) are negligible compared to the offset one without changing the resonant contribution ($\propto\gamma/m^3$).
For example, the ${\rm S1}$ benchmark exhibits an out-of-pole contribution larger than $\gtrsim 0.1\,{\rm cm^2/g}$ in the low-velocity limit, making the velocity dependence incompatible with the stringent constraint from the UFDs.
However, as discussed above, it is possible to reduce the out-of-pole contribution while keeping the resonant contribution unchanged, so that the velocity dependence becomes similar to that of the $p$-wave benchmarks, e.g., ${\rm P1}$.
To be consistent with the constraint from the UFDs, one needs to require a narrow width of a resonance, $\gamma\lesssim 10^{-7}\,(m/{\rm GeV})^{3/2}\,[v_R/(100\,{\rm km/s})]^2$.

The ${\rm P1}$, ${\rm P2}$ and ${\rm P3}$ benchmarks are for $p$-wave resonant self-scattering, and their velocity dependence is consistent with the preferred cross section from the dwarfs/LSBs and the stringent constraint from the UFDs;
these three benchmarks will be the focus of this work.
We remark that while the ${\rm P2}$ and ${\rm P3}$ benchmarks seem to evade the constraints from the UFDs, the high-velocity limit of the SIDM cross section at $\langle v_{\rm rel}\rangle\sim 1000\,{\rm km/s}$, i.e., $\sigma_0/m$, is much smaller than the favored values from galaxy clusters.
The common rSIDM parameters for the $p$-wave benchmarks are $v_R=108\,{\rm km/s}$, $\gamma=10^{-3}$ and $m/S^{1/3}=0.4\,{\rm GeV}$ while the values of $\sigma_0/m$ are $0.1$, $0.03$ and $0.001\,{\rm cm^2/g}$ for ${\rm P}1$, ${\rm P}2$ and ${\rm P}3$ benchmarks, respectively.
Contrary to the $s$-wave scattering, the NWA~[Eq.~\eqref{eq:NWA}] is a good estimation of the resonant contribution even for relatively large couplings $\gamma\sim {\cal O}(1)$.
The out-of-pole contribution in the low-velocity limit is velocity suppressed and thus negligible compared to the considered $\sigma_0/m$'s in the $p$-wave benchmarks;
fixing the resonant peak value as $\langle \sigma v_{\rm rel} \rangle / (m\langle v_{\rm rel} \rangle)\simeq 3\,{\rm cm^2/g}$, the out-of-pole contribution is given by $\sim 10^{-6}\,{\rm cm^2/g}\, [\langle v_{\rm rel}\rangle/(v_R/2)]^4 (\gamma/10^{-3}) [(108\,{\rm km/s})/v_R]$.
One possible contribution for the non-zero $\sigma_0/m$ is the $t$-channel exchange of the resonant mediator, i.e., $\sim \gamma^2/m^3$.
Such a minimal contribution is suppressed by the factor of $\gamma v_R^{3-2L}$ compared to the resonant peak cross section, and thus the $p$-wave benchmarks considered in this work require an additional contribution for the desired $\sigma_0/m$.
For example, in the QCD-like theories of DM, DM can be the dark pseudoscalar mesons where the derivative self-couplings of DM provide the constant $\sigma_0$ and a dark vector meson state can mediate the $p$-wave resonant self-scattering~\cite{Tsai:2020vpi}.
\\

{\bf Gravothermal fluid method.}
We study the evolution of isolated resonant SIDM halos by numerically following the gravothermal fluid equations.
In this method, the system of DM particles is described by a set of fluid conservation equations.
In the collisionless limit, the equations are directly derived by taking the moments of the collisionless Boltzmann equation.
To take into account the effect of self-scattering, the energy conservation equation is modified.
For a general distribution function, the hierarchy of conservation equations is not closed at the finite truncation of the hierarchy.
Nevertheless, as we assume the spherical symmetry of halos with skew-free velocity distribution with isotropic velocity dispersion~[Eq.~\eqref{eq:MBdist}], the closed set of conservation equations is obtained~\cite{Ahn:2004xt}.
We further assume that the gravothermal evolution is quasi-static so that the hydrostatic equilibrium is achieved at each moment~\cite{Balberg:2002ue,Balberg:2001qg,Koda:2011yb,Pollack:2014rja,Essig:2018pzq,Nishikawa:2019lsc}:
\begin{subequations}
\label{eq:gravothermaleqns}
    \noindent\begin{minipage}{0.27\textwidth}
    \begin{equation}
\frac{\partial\left(\rho\nu^{2}\right)}{\partial r}+\frac{GM\rho}{r^{2}}=0\label{eq:1Euler}\,,
\end{equation}
    \end{minipage}%
    \begin{minipage}{0.21\textwidth}
    \begin{equation}
\frac{\partial M}{\partial r}=4\pi r^{2}\rho\,,\label{eq:1mass}
\end{equation}
    \end{minipage}%
    \\
     \begin{minipage}{0.45\textwidth}
\begin{equation}
\frac{3}{\nu}\left(\frac{\partial\nu}{\partial t}\right)_{M}-\frac{1}{\rho}\left(\frac{\partial\rho}{\partial t}\right)_{M}=-\frac{1}{4\pi r^2\rho\nu^2}\frac{\partial L}{\partial r}\label{eq:1entropy}\,,
\end{equation}
    \end{minipage}\vskip1em
\end{subequations}
\noindent where $\rho(r,t)$ and $\nu(r,t)$ are mass density and one-dimensional velocity dispersion, respectively.
$M(r,t)$ is the fluid mass enclosed within radius $r$, and $G$ is the Newton's constant.
$(\partial_t)_M$ is the Lagrangian time derivative which refers to changes within the fluid element as it changes its state and location.
Eq.~\eqref{eq:1Euler} is the condition for hydrostatic equilibrium and Eq.~\eqref{eq:1mass} defines the enclosed fluid mass $M$.

\begin{figure*}[t]
\centering
\includegraphics[width=0.99\textwidth]{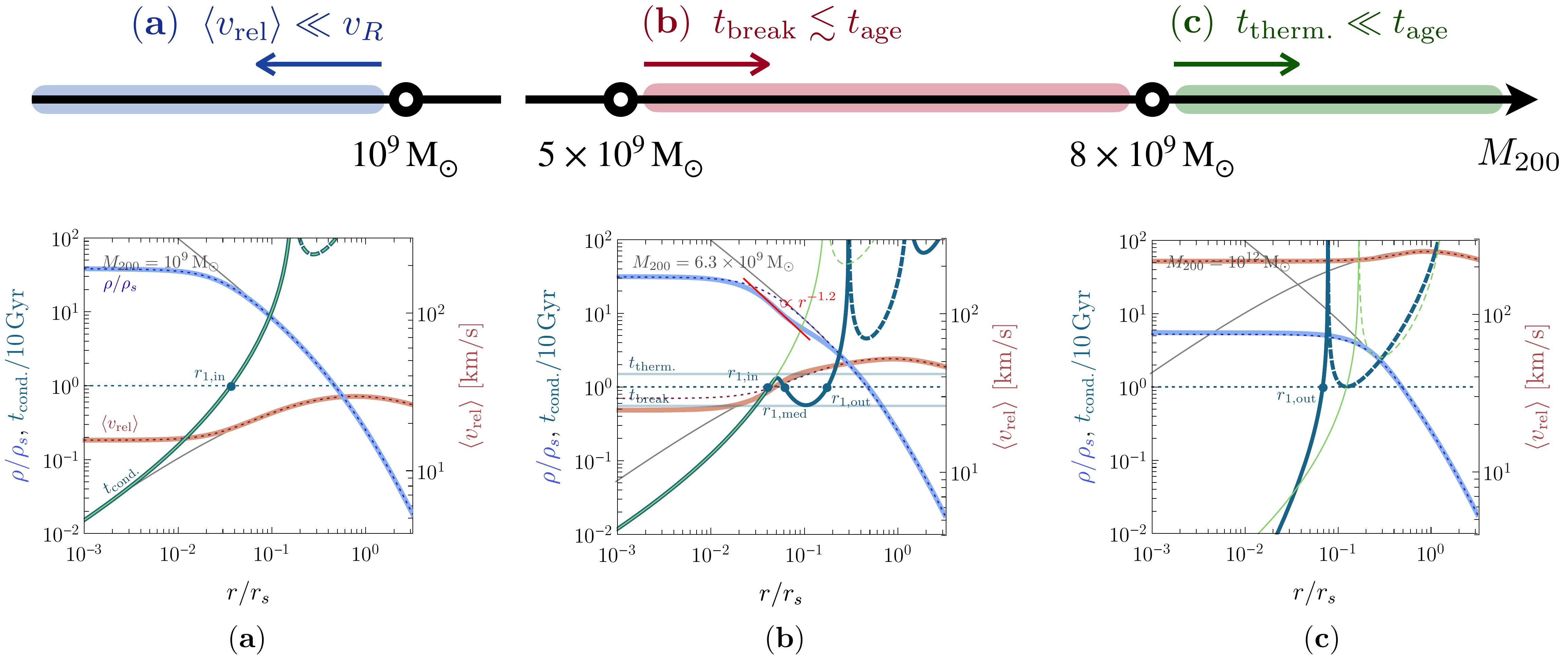} 
\caption{Structure of rSIDM halos for the ${\rm P}2$ benchmark at $t_{\rm age}=10\,{\rm Gyr}$. Blue (orange) curve represents the density ($\langle v_{\rm rel}\rangle$) profile; solid curves represent the rSIDM halos and dotted curves are the corresponding cSIDM profiles with identical inner-most density.
Gray curves are the initial NFW profiles. Green curves are the heat conduction timescale profiles~(see Section~\ref{section:rSIDMevolution} for the definition); solid (dashed) corresponds to net heating (cooling). {\bf (a)} - A halo with DM scattering velocities much smaller than the resonant velocity, $v_R=108\,{\rm km/s}$. The halo structure is the same as that in the cSIDM (dotted) with $\sigma/m=\sigma_0/m=0.03\,{\rm cm^2/g}$. {\bf (b)} - A halo that exhibits a density break at present due to resonant self-scattering of DM; $t_{\rm break}\simeq5.7\,{\rm Gyr}$ and $t_{\rm therm.}\simeq 17\,{\rm Gyr}$~(see Section~\ref{section:rSIDMevolution} for the definition of the timescales). The cSIDM profile is for $\sigma/m=0.035\,{\rm cm^2/g}$ {\bf (c)} - A halo with complete thermalization of the density break at present, i.e., $t_{\rm therm.}\ll t_{\rm age}$. The rSIDM profile is identical to the cSIDM profile with $\sigma/m=0.47\,{\rm cm^2/g}$.}
\label{fig:break}
\end{figure*}

Eq.~\eqref{eq:1entropy} is the energy conservation equation.
The details of microphysics appear in the RHS, which is the rate of heat gain of a fluid element.
$L(r,t)$ is the luminosity which is the power of DM kinetic energy crossing a sphere of radius $r$ by heat conduction;
fluid elements conduct heat with neighboring sites in the radial direction through DM self-interaction.
The heat conduction is modeled by Fourier's law for heat flux:
\begin{equation}
\frac{L}{4\pi r^2}=-\kappa \frac{\partial T}{\partial r}\,,
\end{equation}
where $\kappa$ is the thermal conductivity and $T(r,t)$ is the DM temperature which is related to the one-dimensional velocity dispersion as $\nu=\sqrt{T/m}$.
The general expression for $\kappa$ cannot be derived from the first principles.
Nevertheless, from a simple order-of-magnitude estimation for a DM fluid, the thermal conductivity is estimated as $\kappa \sim \rho/m\times L^2 / T$ where $T$ is the mean time between DM self-scatterings and $L$ is the mean displacement (in radius) that DM particle travels until it deposits/gains its kinetic energy by colliding with another DM particle in a local fluid element.

The estimations of $L$ and $T$ are possible in two limiting regimes: short mean free path (SMFP) and long mean free path (LMFP) regimes.
The two regimes are defined by comparing two length scales, i.e., the free streaming length defined by the self-scattering $\lambda=1/(\rho\sigma/m)$, and the size of halo quantified by the Jeans lengthscale $H=\sqrt{\nu^2/4\pi G\rho}$.
The SMFP regime is the limit of $\lambda\ll H$, where $L$ ($T$) is determined by $\lambda$ [the self-scattering timescale $t_{\rm self}=1/(\rho \langle v_{\rm rel} \rangle\sigma/m)$].
The thermal conductivity in the SMFP regime can be calculated in the Chapman-Enskog expansion~\cite{Outmezguine:2022bhq,Yang:2022hkm}:
\begin{equation}
\kappa_{\rm SMFP}=\frac{75\sqrt{\pi}}{64}\frac{\nu}{\sigma_0 K_5(\nu)},\;\;\; K_p(\nu)=\frac{\langle \sigma v_{\rm rel}^p \rangle}{\sigma_0 \langle v_{\rm rel}^p \rangle}\,.
\end{equation}

The LMFP regime is the opposite limit, i.e., $\lambda \gg H$, where the length separation between two successive scatterings is now determined by the system size $H$;
in this work, as we focus on the core expansion phase of rSIDM halos with $\sigma/m\lesssim {\cal O}(1)\,{\rm cm^2/g}$, the halo evolutions reside in the LMFP regime.
Thermal conductivity in this regime cannot be determined from first principles but can be estimated through a naive dimensional analysis: $\kappa\sim(3C/2)(\rho H^2/m t_{\rm self})$.
The dimensionless constant $C$ is calibrated by comparing with the halo evolution in $N$-body simulations;
as we focus on the halo evolution in the core expansion phase, we adopt $C\simeq0.75$ following \cite{Koda:2011yb,Essig:2018pzq}.
The dimensional analysis for the thermal conductivity shows that $\kappa_{\rm LMFP}\propto \sigma$, but it has not been clarified which distribution-averaged cross section, i.e., $K_p$, is most suitable to describe heat conduction in the LMFP regime;
possible choices may be $\propto \langle \sigma v_{\rm rel}^3\rangle$, acknowledging the fact that the heat conduction rate is proportional to the energy transfer rate~\cite{Outmezguine:2022bhq}, or $\propto \langle \sigma v_{\rm rel}^5\rangle$ by further taking into account the additional velocities appearing in the lengthscale-squared $H^2$~\cite{Yang:2022zkd}.
In this work, we simply assume that $\kappa_{\rm LMFP}\propto K_1$ to better manifest the connection between our analyses and the cross section that is directly read off by comparing the isothermal Jeans modeling and astrophysical observations~(see, e.g., \cite{Kaplinghat:2013xca,Kaplinghat:2015aga,Kamada:2016euw,Valli:2017ktb,Robertson:2020pxj}):
\begin{equation}
\kappa_{\rm LMFP}=\frac{3C}{2(\pi)^{3/2}}\frac{\rho \nu^3 \sigma_0}{Gm^2}K_1(\nu)\,.
\end{equation}
We interpolate the thermal conductivities in the two regimes as $\kappa^{-1}=\kappa_{\rm SMFP}^{-1}+\kappa_{\rm LMFP}^{-1}$.
Possible changes for choosing different values of $p$ are elaborated in Appendix~\ref{appendix:Kp}.

We assume the initial halo density profile to be the Navarro-Frenk-White (NFW) profile~\cite{Navarro:1995iw,Navarro:1996gj}, $\rho_{\rm NFW}=\rho_s/[r/r_s(1+r/r_s)^2]$ where $r_s$ ($\rho_s$) is the NFW scale radius (density).
Unless noted, we take the scale parameters determined by the halo virial mass ($M_{200}$) using the median mass-concentration relation (at present) in cosmological CDM $N$-body simulations~\cite{Dutton:2014xda}:
\begin{equation}
\begin{aligned}
\rho_s&\simeq 0.011\,{\rm M}_{\odot}/{\rm pc}^3\,\left(\frac{10^{10}\,{\rm M}_{\odot}}{M_{200}}\right)^{0.24}\,,\\
r_s&\simeq 3.43\,{\rm kpc}\, \left(\frac{M_{200}}{10^{10}\,{\rm M}_{\odot}}\right)^{0.44}\,.
\end{aligned}
\label{eq:cM}
\end{equation}

\section{Gravothermal evolution of resonant SIDM halos}  \label{section:rSIDMevolution}

In this section, we study the structural evolution of isolated rSIDM halos by numerically solving the gravothermal equations~[Eq.~\eqref{eq:gravothermaleqns}].
We take mainly the ${\rm P}2$ benchmark (and occasionally the ${\rm P}1$ and ${\rm P}3$ benchmarks) and scope the evolutions of halos of various masses. 
For each benchmark, we delineate the halo mass range where the imprint of resonant DM self-interaction can be explicitly seen in present halo structure, i.e., as a break in the density profile.
For halos outside the mass range, the density-profile break disappears through thermalization before present, or does not form from the beginning;
in both cases, their present structure is well approximated by that in cSIDM.
However, even though their present structure is indistinguishable from cSIDM halos, the thermalization of the density-profile break exhibits unique dynamics that may leave an imprint on the kinematic properties of stars of specific age and metallicity, as will be discussed in Section~\ref{section:observations}.
We therefore study how the thermalization dynamics depend on rSIDM parameters by comparing the dynamics among our benchmarks.
\\ 

\begin{figure*}[t]
\centering
\includegraphics[width=0.75\textwidth]{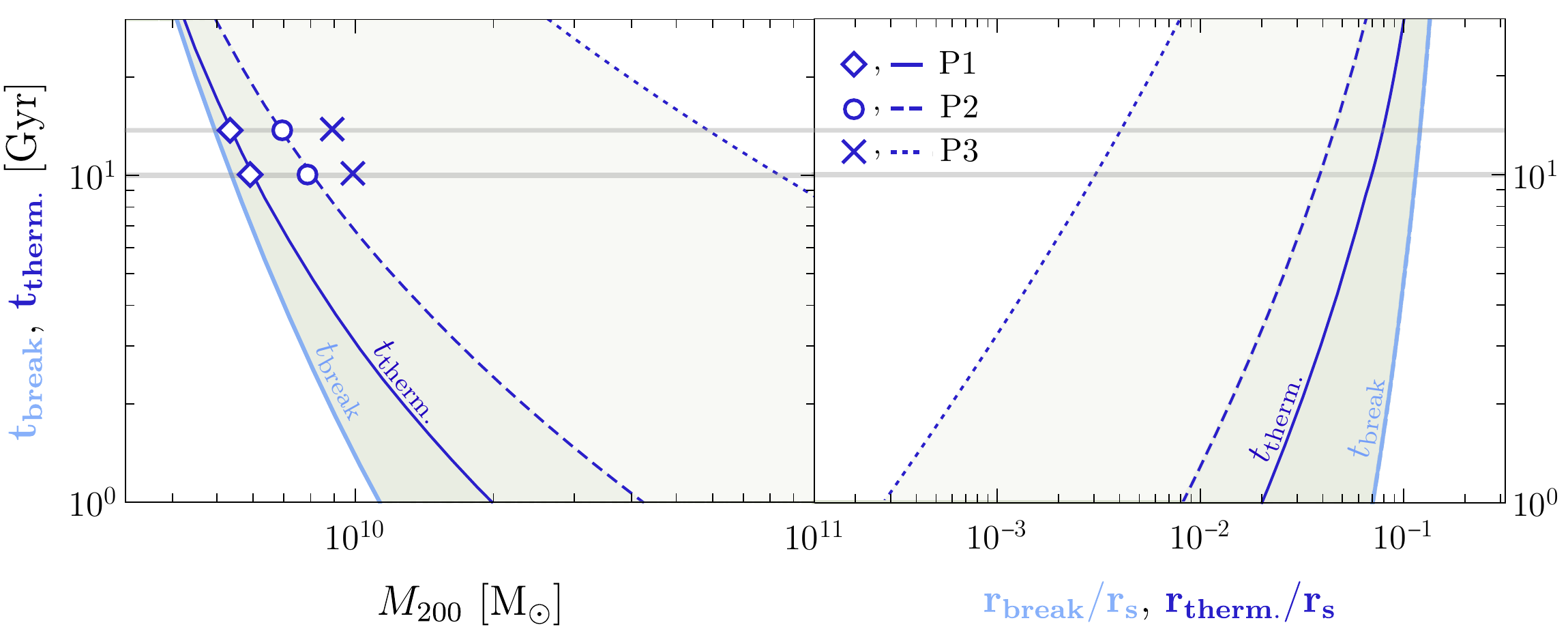} 
\caption{{\bf (Left)} - The halo-mass range where the density break from rSIDM may be probed at present. The light-blue (blue) curves represent the halo-mass dependence of $t_{\rm break}$ ($t_{\rm therm.}$). The horizontal lines represent $t_{\rm age}=10\,{\rm Gyr},\,13.8\,{\rm Gyr}$.  Given a halo age, our expected mass range is determined by the condition $t_{\rm break}\lesssim t_{\rm age}\lesssim t_{\rm therm.}$. The expected high(low)-end of the mass range is determined by the condition $t_{\rm age}=t_{\rm therm.}$ ($t_{\rm age}=t_{\rm break}$). The plot markers are the actual high-end of the mass range from our simulations. While our expected mass range coincides well with the simulations for ${\rm P1}$ ($\sigma_0/m=0.1\,{\rm cm^2/g}$) and ${\rm P2}$ ($\sigma_0/m=0.01\,{\rm cm^2/g}$) benchmarks~(see also the left panel of Fig.~\ref{fig:thermalization} for the ${\rm P2}$ benchmark), the high-end of the mass range is overestimated in the ${\rm P3}$ ($\sigma_0/m=0.001\,{\rm cm^2/g}$) benchmark.  {\bf (Right)} - The expected range in the halo radius where the density break is located. The horizontal axis $r_{\rm break}/r_s$ ($r_{\rm therm.}/r_s$) is the expected radius at which the density break starts to form (thermalize) at $t_{\rm age}=t_{\rm break}$ ($t_{\rm age}=t_{\rm therm.}$).}
\label{fig:range}
\end{figure*}

{\bf Halo-mass dependence.}
For halos with the maximal DM velocity far below the resonant velocity, the DM particles miss the resonance and thus their evolution proceeds as in cSIDM with the offset cross section $\sigma_0/m$;
as shown in the panel ${\rm (a)}$ of Fig.~\ref{fig:break}, the rSIDM (solid) and the cSIDM (dotted) density (blue) and velocity dispersion (orange) profiles are identical.
The heat conduction timescale $t_{\rm cond.}(r)$ is a convenient measure to identify the region inside a halo significantly affected by DM self-interaction;
$t_{\rm cond.}$ is defined as the inverse of the RHS of Eq.~\eqref{eq:1entropy} times an ${\cal O}(1)$ fudge factor, assuming the initial NFW density profile and the corresponding velocity dispersion profile~\cite{Lokas:2000mu}.
We choose the fudge factor to be $\sim 1/3$, as will be discussed soon.
For a region with $t>t_{\rm cond.}$, we may expect a significant deviation from the initial NFW profile.
The $t_{\rm cond.}$ profile for the rSIDM (cSIDM) is shown as a green (light green) curve in Fig.~\ref{fig:break}, where the solid curve corresponds to the region of net heating, and the dashed curve corresponds to net cooling;
in the core expansion phase, the heat conduction from DM self-scattering works in a way to transfer heat from the region of $r/r_s\sim1$ to smaller radii.
The two $t_{\rm cond.}$ distributions, i.e., for the rSIDM and the cSIDM, are identical, showing that the evolution of halos far below the resonant velocity is basically identical to that in cSIDM with $\sigma=\sigma_0$.
The size of the central core is conveniently predicted by comparing a given time $t$ with $t_{\rm cond.}$;
for the region of $t>t_{\rm cond.}$, the heat conduction is efficient enough to thermalize DM and thus form a uniform core.
We define the radius at which $t=t_{\rm cond.}$ as $r_{\rm 1,in}$ if the conduction timescale is determined by the offset cross section;
the subscript ``in" will become clear in the following discussion.
The fudge factor for $t_{\rm cond.}$ is chosen so that $r_{\rm 1,in}$ roughly coincides with the core size.

For halos of the maximal DM velocity close to the resonant velocity, only a localized region inside a halo may feel the resonance.
Such a case is shown in panel ${\rm (b)}$ of Fig.~\ref{fig:break}.
Contrary to the panel ${\rm (a)}$ where the $t_{\rm cond.}$ distribution is a monotonically increasing function with radius, $t_{\rm cond.}$ in panel ${\rm (b)}$ exhibits a local minimum at a certain radius;
the vicinity of such a radius is the region where the heat conduction is enhanced due to the resonant self-scattering.
For our benchmarks, such a local minimum can be defined for halos whose DM scattering velocities overlap with the resonant velocity range at some radius, i.e., $0.5 \lesssim \langle v_{\rm rel} \rangle/v_R \lesssim 1.5$.
We define the $t_{\rm cond.}$ (radius) at such a local minimum as $t_{\rm break}$ ($r_{\rm break}$).
Shortly after $t\gtrsim t_{\rm break}$, there exist three radii where $t=t_{\rm cond.}$;
towards larger radius, we define them as $r_{\rm 1,in}$, $r_{\rm 1, med}$, and $r_{\rm 1,out}$, respectively.
$r_{\rm 1,in}$ is determined by the offset cross section, as in the panel ${\rm (a)}$ of Fig.~\ref{fig:break}.
$r_{\rm 1, med}$ and $r_{\rm 1,out}$ are determined by the resonant cross section.
Contrary to panel $({\rm a})$, there is a region beyond $r\gtrsim r_{\rm 1,in}$ where the halo profile significantly deviates from the NFW profile due to the heat conduction, i.e., $r_{\rm 1, med} \lesssim r \lesssim r_{\rm 1, out}$.
The separation of the two regions renders a break in the density profile.
Such a break in the density profile may be probed if it occurs at the radius relevant to astrophysical observations, e.g., rotation curves or stellar LOSVD profiles.
The generation of the break is solely due to the resonant self-scattering, and can serve as a smoking-gun signature of rSIDM.
In order for the density-profile break to appear before the present, the halo age should satisfy $t_{\rm age}\gtrsim t_{\rm break}$.
For future convenience, let us define the logarithmic slope of the density-profile break $\alpha_{\rm break}$, which is defined as the steepest slope for $r\lesssim r_{\rm 1,med}$;
in panel $({\rm b})$ of Fig.~\ref{fig:break}, $\alpha_{\rm break}\simeq -1.2$.
To be concrete, we define the position of the density break as the point of local minimum of $\alpha_{\rm break}<-1$.

\begin{figure*}[t]
\centering
\includegraphics[width=1\textwidth]{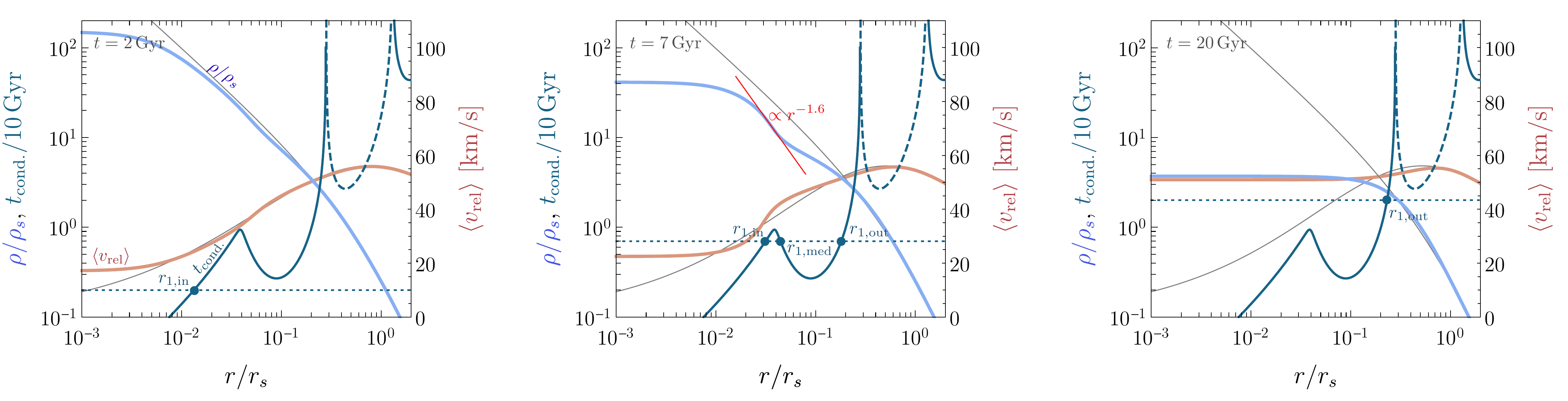} 
\caption{
Time evolution of an rSIDM halo in the ${\rm P2}$ benchmark. The halo mass is $8\times 10^{9}\,{\rm M_\odot}$ with median concentration. {\bf (Left)} - A snapshot of the halo profile at $t \sim t_{\rm break}$. A density break starts to appear around $r_{\rm break}$. {\bf (Middle)} - Halo profile for $t_{\rm break} \lesssim t \lesssim t_{\rm therm.}$. During this period, the density break manifests itself by increasing its slope; in the plot, $\alpha_{\rm break} \simeq -1.6$. {\bf (Right)} - Halo profile for $t > t_{\rm therm.}$. The density break has disappeared due to the thermalization of the core.  
}
\label{fig:evolution}
\end{figure*}

Eventually, the density break thermalizes to disappear.
The dynamics of the thermalization will be discussed in detail in the next subsection.
Similarly to $t_{\rm break}$, one may attempt to estimate the time when the thermalization completes through the $t_{\rm cond.}$ profile.
We define $t_{\rm therm.}$ as the time when the two regions of efficient heat conduction, i.e., $r\lesssim r_{\rm 1,in}$ and $r_{\rm 1, med} \lesssim r \lesssim r_{\rm 1, out}$, merge into one, i.e., $t_{\rm therm.}$ is the local maximum of the $t_{\rm cond.}$ profile; we define the radius at $t_{\rm cond.}=t_{\rm therm.}$ as $r_{\rm therm.}$.
We remark that $t_{\rm therm.}$, which is defined by the NFW profile, tends not to be a good estimation of the actual thermalization time for more visible development of the density break.
This is because at the time of $t_{\rm therm.}$, the density break can be significantly developed so that the global halo profile significantly deviates from the initial NFW profile.
Nevertheless, we will use $t_{\rm therm.}$ as a rough indicator for the thermalization time.
As we consider halos with the maximal DM velocity much larger than the resonant velocity, the region of resonant velocities is located at smaller radii ($r/r_s\ll 1$) and $t_{\rm therm.}$ becomes shorter than $t_{\rm age}$;
the density break forms in a smaller radius and thermalizes before the present.
Panel $({\rm c})$ of Fig.~\ref{fig:break} represents such a halo.
There, we confirm that the resultant density profile after the thermalization is virtually indistinguishable from a cSIDM profile with a similar central density.

We now delineate the halo-mass range where the imprint of resonant self-scattering is explicit (as a density break) in their present structures.
We estimate the range by requiring $t_{\rm break}\lesssim t_{\rm age}\lesssim t_{\rm therm.}$.
The estimations for the $p$-wave benchmarks are shown in Fig.~\ref{fig:range};
the left panel shows the estimated halo-mass range depending on the assumed halo age, and the right panel shows the corresponding range in radius where the density break can be manifested.
The low-end of the halo-mass range, which is determined by $t_{\rm break}=t_{\rm age}$ (light blue curve), are identical among the $p$-wave benchmarks since they exhibit the same resonant self-scattering.
The high-end of the range is larger for smaller $\sigma_0/m$ since $t_{\rm therm}$ (blue curves) is longer for smaller $\sigma_0/m$.
We confirm that the range indeed increases as we consider smaller $\sigma_0/m$, as shown as the plot markers in Fig.~\ref{fig:range}.
However, our estimation tends to overpredict the high-end of the halo-mass range as we consider smaller offset cross section.
This is because smaller values of $\sigma_0/m$ result in density breaks with larger $|\alpha_{\rm break}|$ where the global halo profiles deviate more from the NFW profile, as will be discussed in the next subsection.
\\

\begin{figure*}[t]
\centering
\includegraphics[width=0.45\textwidth]{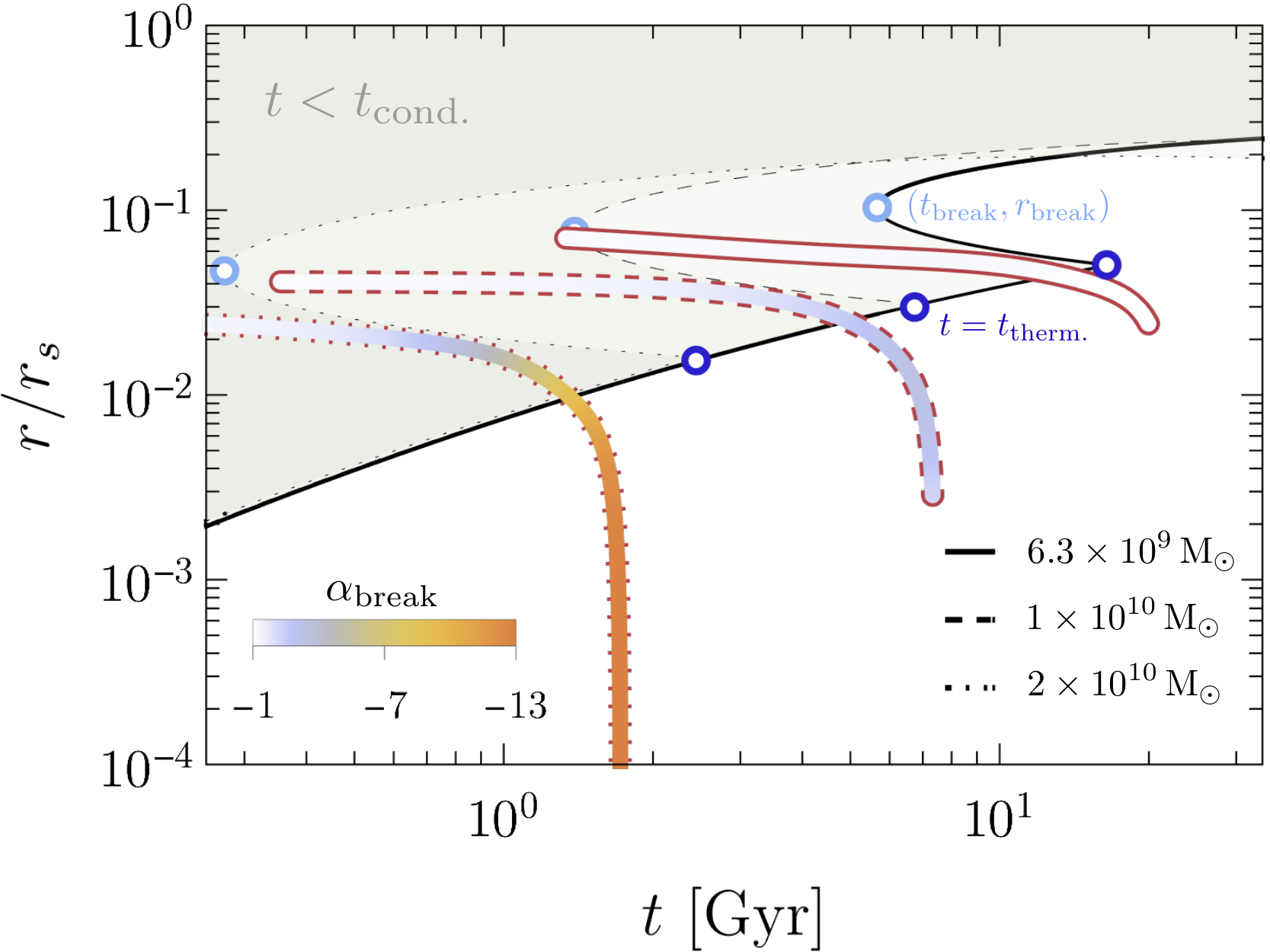} \hspace{1cm}
\includegraphics[width=0.45\textwidth]{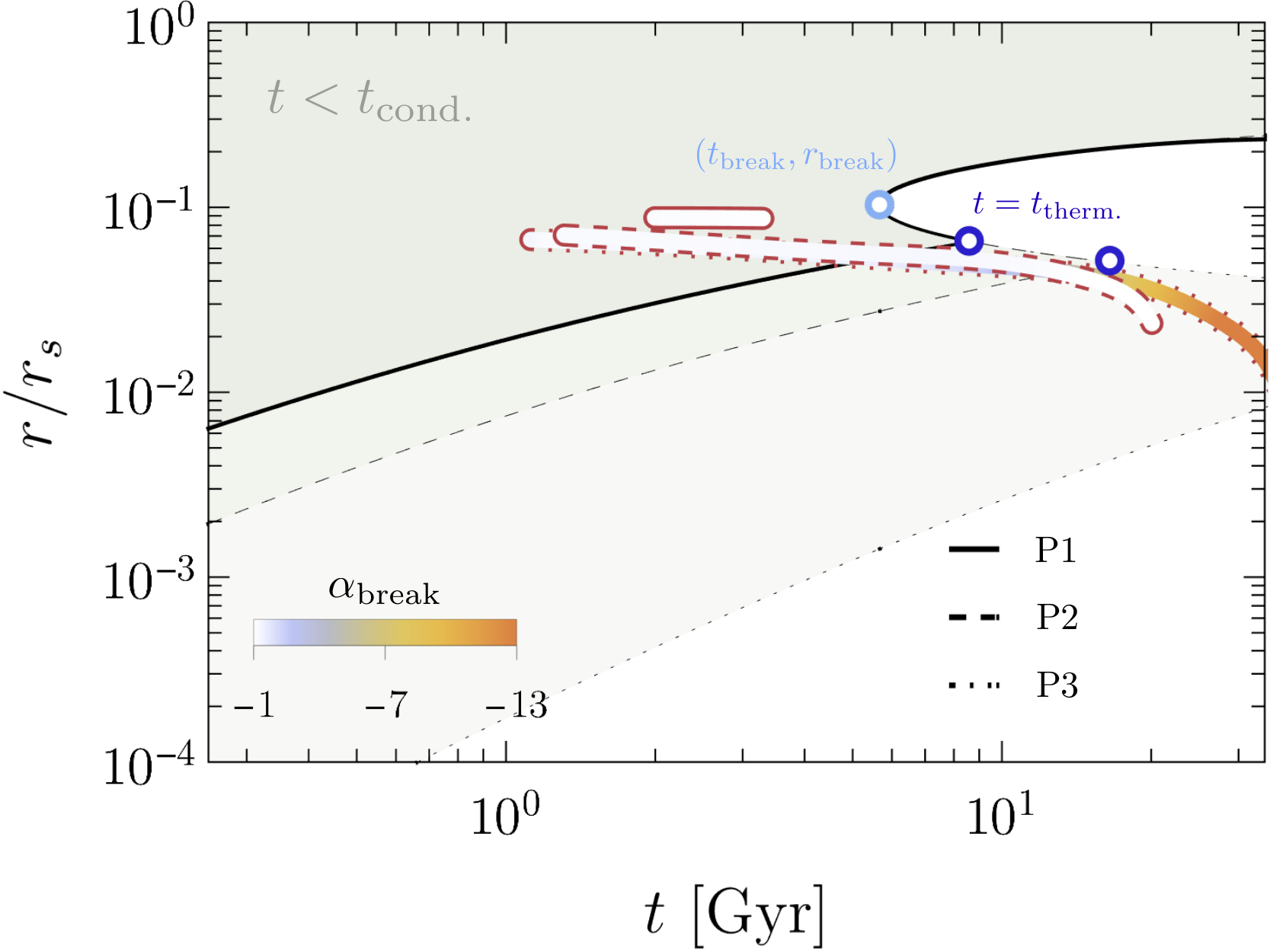} 
\caption{{\bf (Left)} - Evolution of the density break for the ${\rm P2}$ ($\sigma_0/m=0.01\,{\rm cm^2/g}$) benchmark in various halo masses. The colored bars enclosed by red curve represent the position of the density break at a given time; the color scheme shows the evolution of the logarithmic slope of the density break. For the halo of mass $2\times 10^{10}\,{\rm M_\odot}$, the density break thermalizes by propagating towards the origin completely; the density break becomes steeper as it propagates. For the halo masses of  $6.3 \times 10^{9}\,{\rm M_\odot}$ and $10^{10}\,{\rm M_\odot}$, the density break thermalizes by relaxing the slope at a finite radius. The colored region enclosed by black curves represents the radii where the heat conduction is inefficient, i.e., $t<t_{\rm cond.}$.
{\bf (Right)} - Same as the left panel but in the ${\rm P1}$ ($\sigma_0/m=0.1\,{\rm cm^2/g}$), ${\rm P2}$ ($\sigma_0/m=0.01\,{\rm cm^2/g}$), and ${\rm P3}$ ($\sigma_0/m=0.001\,{\rm cm^2/g}$) benchmarks for a given halo mass, $M_{200}=6.3\times 10^{9}\,{\rm M_\odot}$.
}
\label{fig:thermalization}
\end{figure*}

{\bf Thermalization dynamics.}
For halos whose maximal DM scattering velocity is larger than $0.5\,v_R \lesssim \langle v_{\rm rel} \rangle$, a density break forms and eventually thermalizes.
The time evolution of such halos is described in Fig.~\ref{fig:evolution}.
The mass of the presented halo is $8 \times 10^9\,{\rm M_\odot}$ with which the density break starts to thermalize around $t_{\rm age}=10\,{\rm Gyr}$, according to Fig.~\ref{fig:range}.
The left panel is the time when the density break starts to form, i.e., $t \sim t_{\rm break}$.
As the halo evolves further, the density break manifests itself by increasing its logarithmic slope $|\alpha_{\rm break}|$.
The density break is located within the region $r_{\rm 1,in} \lesssim r \lesssim r_{\rm 1,med}$ where we expect the heat conduction is not efficient enough to thermalize the region~(see the middle panel of Fig.~\ref{fig:evolution}).
The density break connects two regions of distinct DM densities, i.e., $r\lesssim r_{\rm 1, in}$ and $r_{\rm 1,med} \lesssim r \lesssim r_{\rm 1,out}$.
The DM density of the former region is determined by $\sigma_0/m$, and that of the latter region is determined by the resonant self-scattering.

Since the region $r_{\rm 1,in} \lesssim r \lesssim r_{\rm 1,med}$ shrinks with time, we can expect the slope of the density profile to increase until $t \lesssim {\cal O}(t_{\rm therm})$.
Note that larger $|\alpha_{\rm break}|$ renders larger heat conduction rate around the density break.
Such an enhanced heat conduction rate is much larger than one can expect in the initial NFW profile.
Therefore, for halo profiles with large $|\alpha_{\rm break}|\gg 1$, $t_{\rm therm}$ loses its significance as an approximate thermalization time of the density break;
this is why the estimated high-end of the halo-mass range dramatically fails to predict the actual value from simulations, as we have discussed in Fig.~\ref{fig:range}.
Nevertheless, the density break thermalizes to disappear by the time $t={\cal O} (t_{\rm therm})$.
As shown in the right panel of Fig.~\ref{fig:evolution}, heat conduction is expected to be efficient for $r\lesssim r_{\rm 1, out}$ and thus a core with uniform DM density and velocity dispersion is expected to form.
The thermalization happens through the propagation of the density break towards the halo center.
At the propagation front, the DM fluid is heated to expand in volume which leaves behind a low-density core as in the right panel of Fig.~\ref{fig:evolution}.
We remark that the propagation of the density break induces an abrupt change in the central gravitational potential, affecting the orbits of stars around the thermalization time, as will be discussed in the next section.

Our simulation shows that among the $p$-wave benchmarks, the density break tends to evolve up to larger values of $|\alpha_{\rm break}|$ and complete its propagation all the way down to the origin for smaller values of $\sigma_0/m$.
This is demonstrated in the right panel of Fig.~\ref{fig:thermalization} which shows the evolution of the position and the slope ($\alpha_{\rm break}$) of the density break for a given halo mass;
since ${\rm P3}$ (dotted) benchmark has the smallest offset cross section, $|\alpha_{\rm break}|$ evolves up to larger values compared to ${\rm P1}$ (solid) and ${\rm P2}$ (dashed).
At the same time, the density break propagates relatively further towards the center for smaller values of $\sigma_0/m$.
The evolution of the density break also depends on the halo mass; see the left panel of Fig.~\ref{fig:thermalization}.
For a fixed $\sigma_0/m$, the density break evolves up to larger $|\alpha_{\rm break}|$ and propagates to smaller values of $r/r_s$ in larger halos.
Both the offset cross section and halo mass controls the initial distinction between (the DM densities of) the two regions $r\lesssim r_{\rm 1,in}$ and $r_{\rm 1,med} \lesssim r \lesssim r_{\rm 1,out}$, i.e., the ratio of $r_{\rm 1,in}$ to $r_{\rm 1,med}=r_{\rm 1,out}$ at $t=t_{\rm break}$.
Smaller $\sigma_0/m$ and larger halo mass render larger values of the ratio, as demonstrated in Fig.~\ref{fig:thermalization}.

\section{Resonant SIDM in astrophysical observations}
\label{section:observations}

\begin{figure*}[t]
\centering
\includegraphics[width=0.336\textwidth]{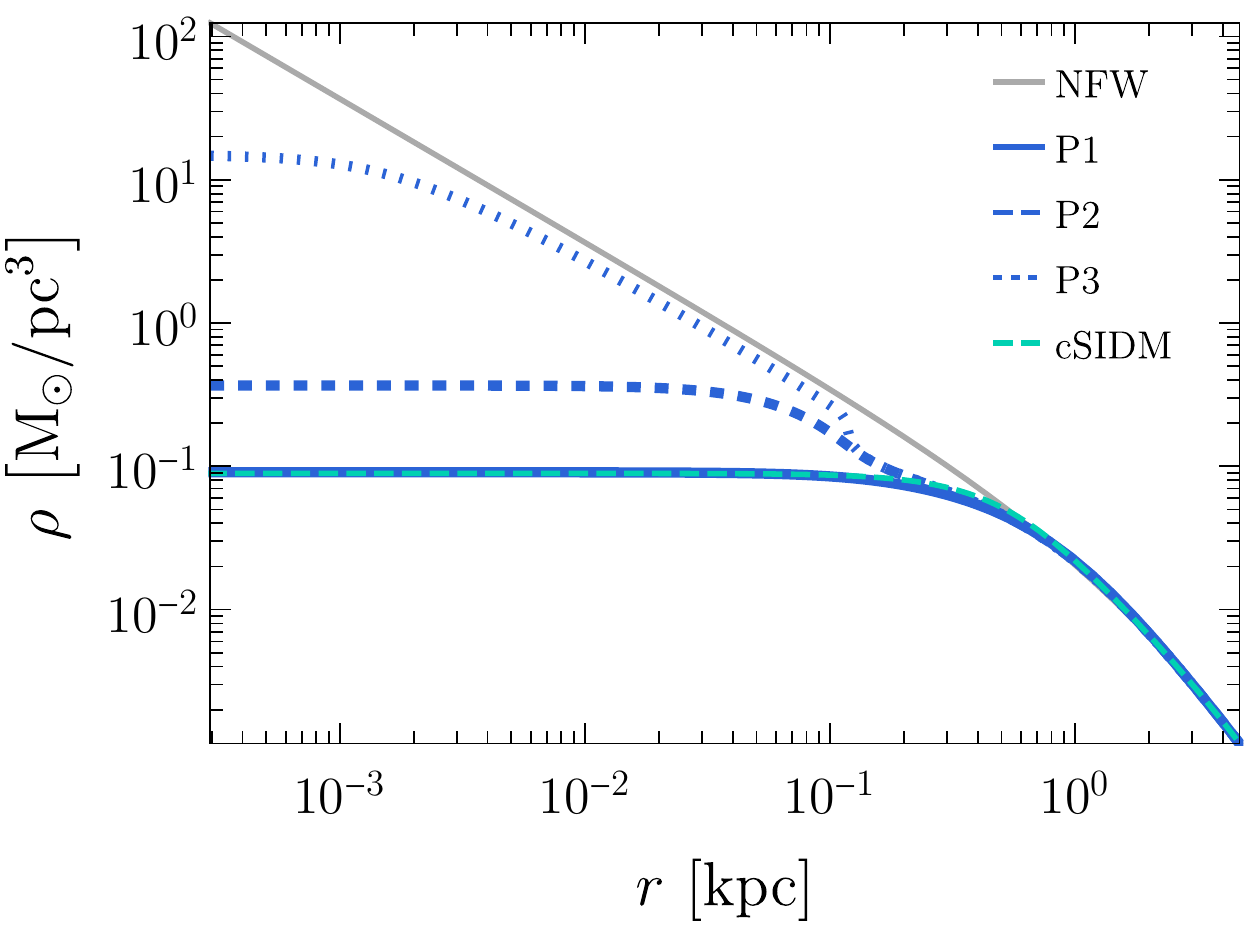}\hspace{0.5cm}
\includegraphics[width=0.322\textwidth]{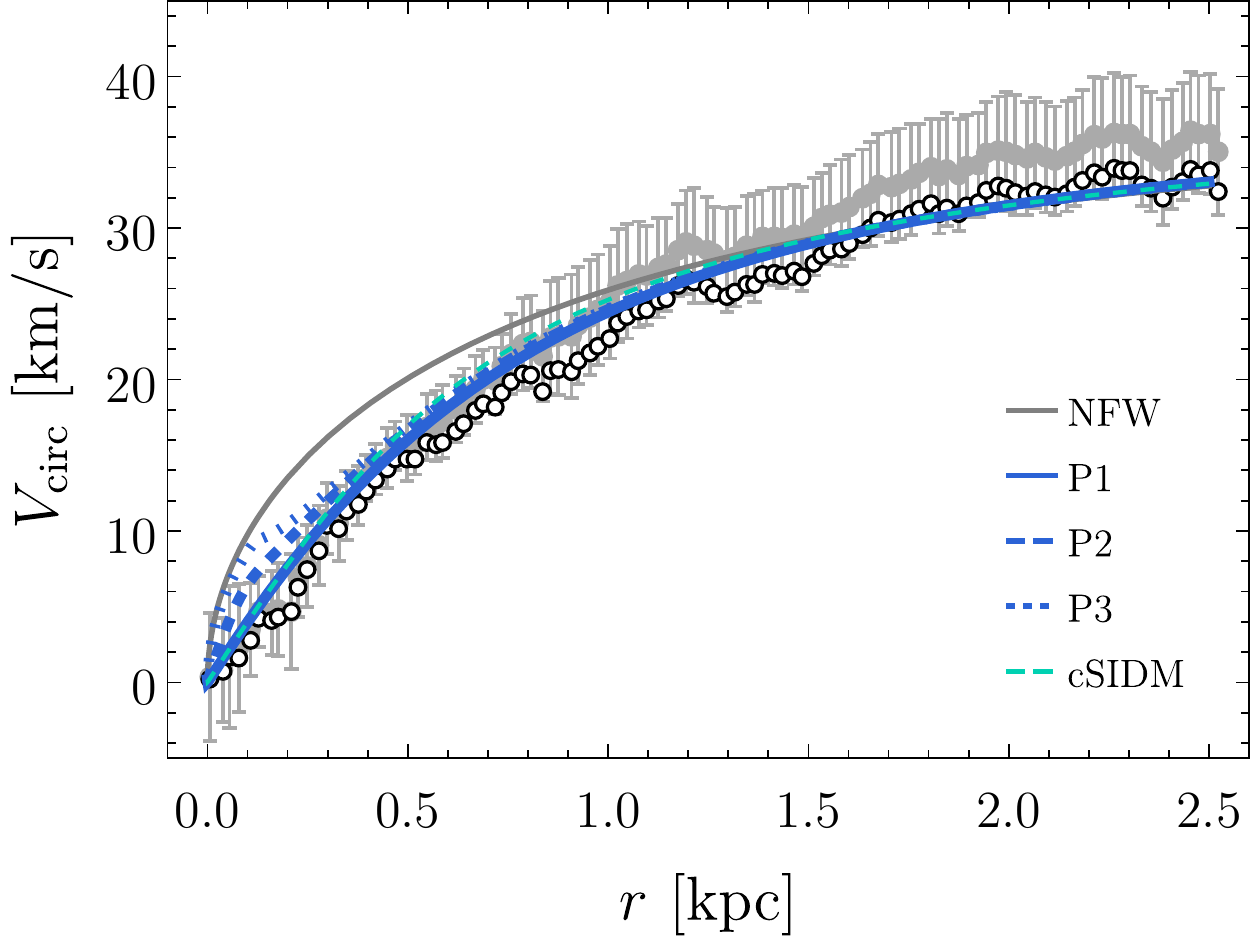} \\\hspace{0.15cm}
\includegraphics[width=0.32\textwidth]{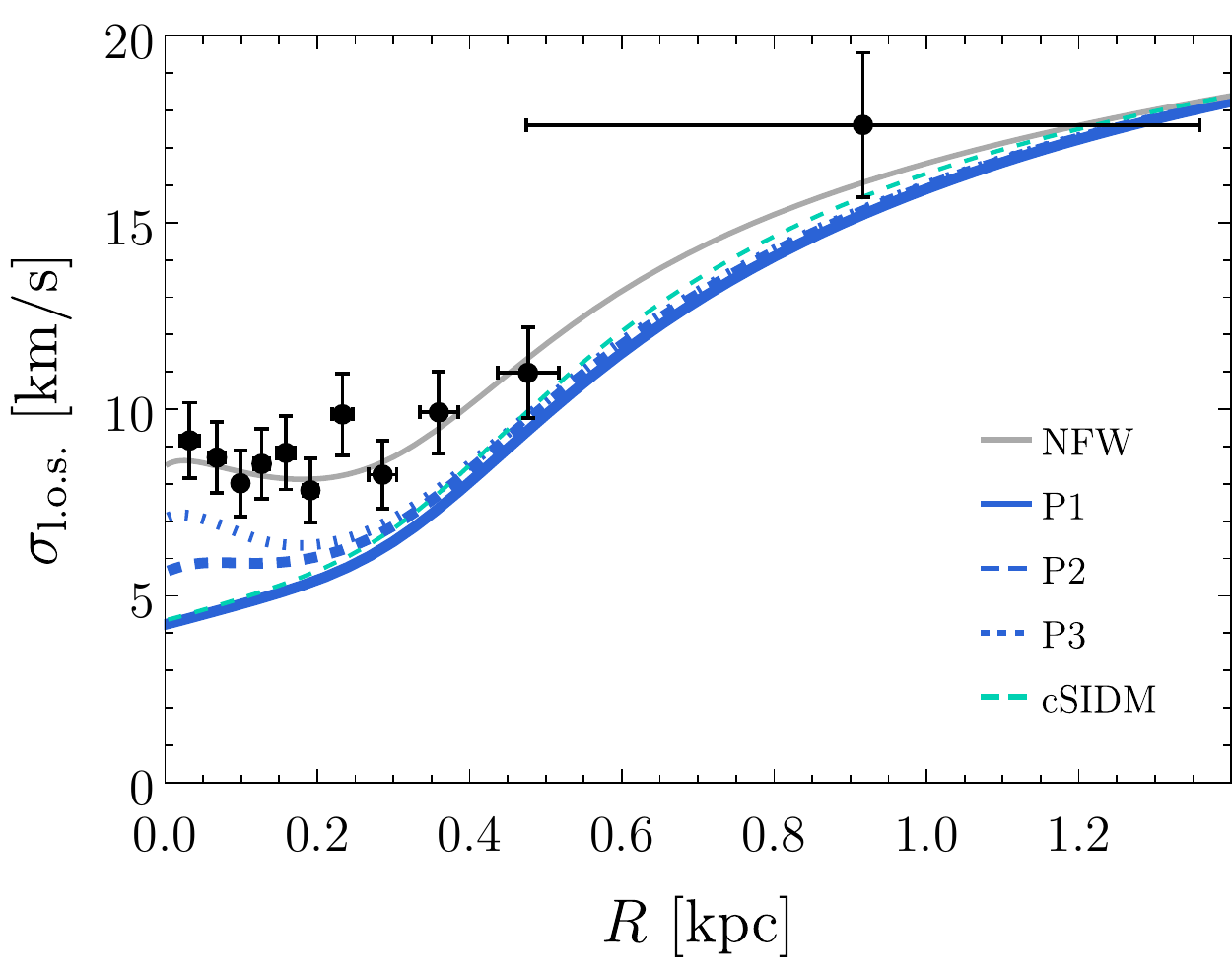} \hspace{0.5cm}
\includegraphics[width=0.32\textwidth]{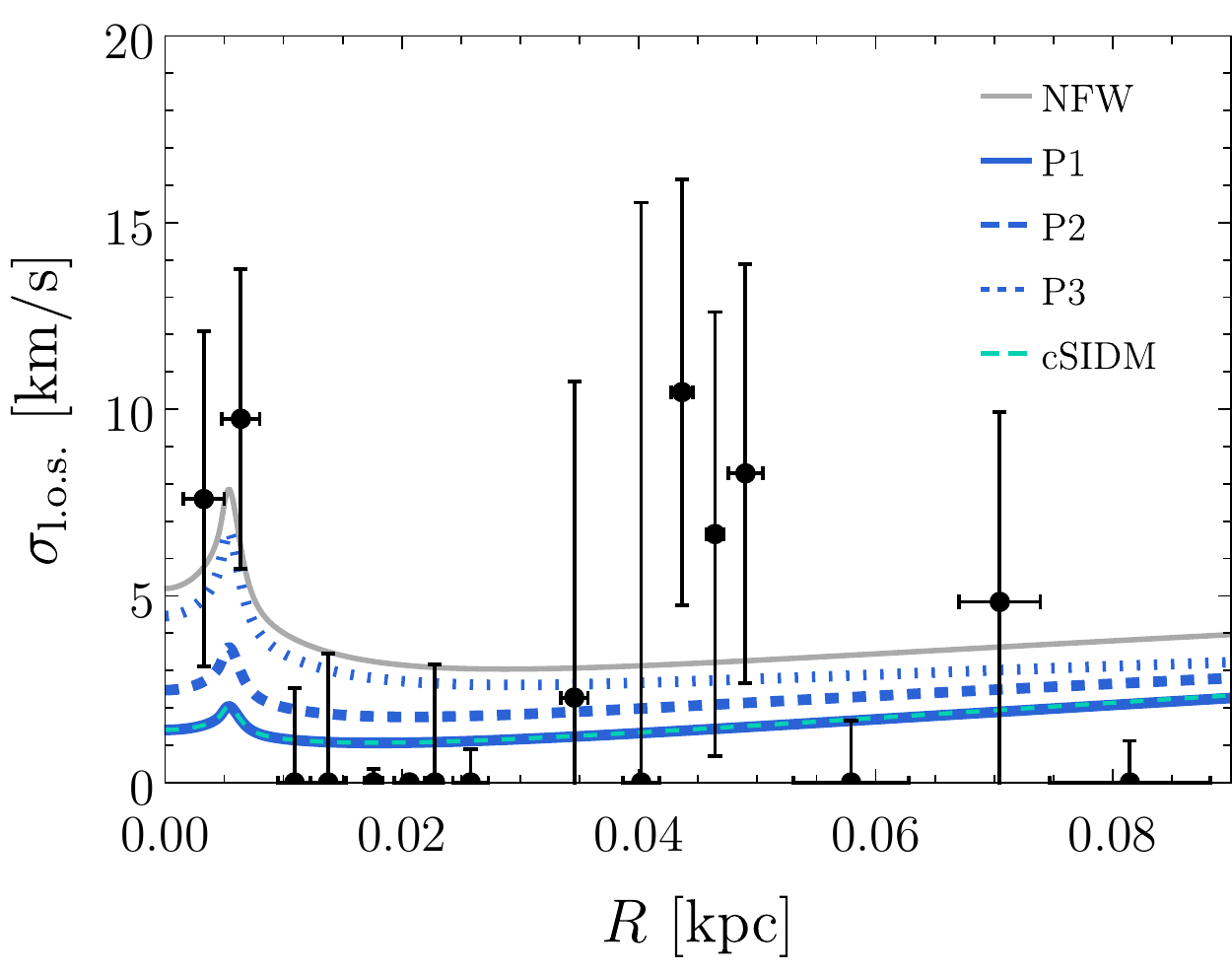} 
\caption{{\bf (Top)} {\it (Left)} - Density profiles of a halo of mass $7\times 10^9\,{\rm M_\odot}$ at $t_{\rm age}=10\,{\rm Gyr}$ in the rSIDM benchmarks (blue); {\rm P1} (solid, $\sigma_0/m=0.1\,{\rm cm^2/g}$), {\rm P2} (dashed, $\sigma_0/m=0.03\,{\rm cm^2/g}$), and {\rm P3} (dotted, $\sigma_0/m=0.001\,{\rm cm^2/g}$). A cSIDM profile (green) that exhibits the same central density with the {\rm P1} benchmark is presented for comparison; the corresponding constant SIDM cross section is $0.33\,{\rm cm^2/g}$. {\it (Right)} - Rotation curve of the halo in the top-left panel. The black curves corresponds to the NFW profile and the blue curves corresponds to the predictions in the $p$-wave benchmarks. The maximal circular velocity is $V_{\rm max}\simeq 36\,{\rm km/s}$. The gray data points represents the observed rotation curve of the WLM galaxy, which has a similar $V_{\rm max}$~\cite{Oh:2015xoa}; we display the data up to $r=2.5\,{\rm kpc}$. The white circles are the inferred contribution of DM to the rotation curve; the error bars are expected to be similar to that of the gray data points.
{\bf (Bottom)} {\it (Left)} - Line-of-sight velocity dispersion (LOSVD) profile of a Draco-like galaxy; halo mass and concentration is the same as the top-left panel. The assumed stellar kinematic parameters are the best-fit values assuming the NFW profile; $(\beta_0, \beta_\infty, r_\beta, \eta, r_{1/2})=(-0.105,-60.7,850\,{\rm pc}, 4.65, 214\,{\rm pc})$~\cite{Note1}. The gray curve corresponds to the NFW profile, and the blue curves are the prediction of the $p$-wave benchmarks. The black data points are the observed LOSVD of Draco~\cite{Note2}; the observed LOSVD is presented as a reference, and we do not attempt to fit the data points in rSIDM.
{\it (Right)} - Same as the bottom-left panel but for a Segue1-like galaxy. The assumed stellar kinematic parameters are the best-fit values assuming the NFW profile; $(\beta_0, \beta_\infty, r_\beta, \eta, r_{1/2})=(-68.0,-0.894,4.42\,{\rm pc}, 10, 23.5\,{\rm pc})$~\cite{Note1}. The black data points are the observed LOSVD of Segue1~\cite{Note2}.
}
\label{fig:probes}
\end{figure*}

In this section, we demonstrate how rSIDM halos look in astrophysical observations, e.g., rotation curves and LOSVD profiles.
The smoking-gun signature of rSIDM is the density-profile break, which is expected to be present for halos in the specific mass range~(see the left panel of Fig.~\ref{fig:range}).
For a concrete demonstration, we focus on a halo of the mass $\sim7\times 10^9\,{\rm M_\odot}$ with median concentration;
according to our results, such a halo would exhibit a density break at present in the ${\rm P2}$ and ${\rm P3}$ benchmarks while the density break is already thermalized for the ${\rm P1}$ benchmark.
We evolve the initial NFW halo for $t_{\rm age}=10\,{\rm Gyr}$ in each benchmark.
In order to infer the density break, the observational probes should be able to resolve the region $0.01 \lesssim r/r_s\lesssim 0.1$~(see the right panel of Fig.~\ref{fig:range} for the estimation on the possible position of the density break);
for the considered halo mass, $r_s\simeq 3\,{\rm kpc}$ and the actual position of the density break is $\sim 0.12\,{\rm kpc}$ in the ${\rm P2}$ and ${\rm P3}$ benchmarks;
see the top-left panel of Fig.~\ref{fig:probes}.
The density profile for the ${\rm P1}$ benchmark is virtually indistinguishable from the cSIDM profile (green) with $\sigma/m=0.33\,{\rm cm^2/g}$.
Note that we do not try to infer the density break or exclude rSIDM benchmarks from the observations.
Nevertheless, we select a couple of galaxy samples that resolve the relevant region and present them alongside with the rSIDM predictions as a reference.

Gas-rich galaxies can provide rotation curve data which directly probes the enclosed mass profile;
$V_{\rm circ}(r)=\sqrt{G M(r)/r}$.
The density-profile break is seen as a break in the central rotation curve;
this is shown in the top-right panel of Fig.~\ref{fig:probes}.
The blue curves are the resultant rotation curves of the halo.
For the ${\rm P1}$ benchmark, the inner rotation curve scales linearly with radius, reflecting the existence of a uniform density core.
For the other benchmarks, there is a transition of the rotation curve around the position of the density break;
rotation curves are NFW-like (black) inside the density break, and they converge to that of the ${\rm P1}$ at larger radius.
As a reference, we display the observed rotation curve of the WLM galaxy selected from the LITTLE THINGS samples~\cite{Oh:2015xoa};
the sample has similar maximal circular velocity to our halo and data is available down to very small radii $\lesssim 0.01\,{\rm kpc}$.
If WLM's mass and concentration were similar to that of our halo, the ${\rm P3}$ benchmark would be disfavored.

Contrary to the Local Group dwarfs like WLM, MW's satellite galaxies are gas-poor and lacks the rotational feature.
For such systems, stellar kinematic data is used to infer the density profile;
for most galaxies, only the line-of-sight motions of stars are available.
The standard quantity to infer the DM distribution is the LOSVD profile.
For spherical and DM-dominated galaxies, we get the following Jeans equation by integrating the collsionless Boltzmann equation for stellar velocity distribution~\cite{GalacticDynamics}:
\begin{equation}
\frac{\partial (n_\star \sigma^2_r)}{\partial r} + \frac{2 \beta n_\star \sigma_r^2}{r} = -\frac{G M n_\star}{r^2}\,,
\end{equation}
where $n_\star(r)$ stellar number density and $\sigma_r(r)$ is their radial velocity dispersion.
The velocity dispersion of stars satisfy $\sigma_\theta(r)=\sigma_\phi(r)$ from spherical symmetry.
The anisotropy parameter of stellar orbits is defined as $\beta(r)=1-\sigma_\theta^2/\sigma_r^2$.
Following Ref.~\cite{Baes:2007tx,Hayashi:2020syu}, we parametrize the anisotropy parameter as
\begin{equation}
\beta(r) = \frac{\beta_0+\beta_\infty(r/r_\beta)^\eta}{1+(r/r_\beta)^\eta}\,,
\end{equation}
where $\beta_0$ ($\beta_\infty$) is the inner (outer) anisotropy, $r_\beta$ is the transition radius, and $\eta$ is the sharpness of the transition.
Given the stellar dispersion profile, the projected LOSVD is given by
\begin{equation}
\sigma_{\rm l.o.s.}^2(R)=\frac{2}{\Sigma(R)}\int^\infty_R dr \left[1-\beta(r)\frac{R^2}{r^2}\right] \frac{n_\star(r)\sigma_r^2(r)}{\sqrt{1-R^2/r^2}}\,,
\end{equation}
where $R$ is the projected radius and $\Sigma(R)$ is the projected surface density of stars.
In order to solve the Jeans equation, we use the enclosed mass profile of our simulated halo, and use the Plummer profile~\cite{GalacticDynamics} for $n_\star$ with the observed half light radius $r_{1/2}$.
We adopt the best-fit stellar kinematic parameter sets of Draco and Segue 1 for demonstration;
they are known to be one of the most cuspy classical dwarfs and UFDs, respectively (see, e.g., Refs.~\cite{Read:2018pft,Hayashi:2020jze,Hayashi:2022wnw}).
We choose the two galaxies because their best-fit NFW parameters is similar to our considered halo;
the best-fit NFW parameters $(\rho_s/({\rm M_\odot/pc^3}), r_s/{\rm kpc})$ to their observed LOSVD data are $(0.014,2.9)$ for Draco and $(0.032,2.4)$ for Segue 1, while the parameters for the initial NFW profile of our simulated halo are $(0.012,3.0)$.

\begin{figure*}[t]
\centering
\includegraphics[width=0.7\textwidth]{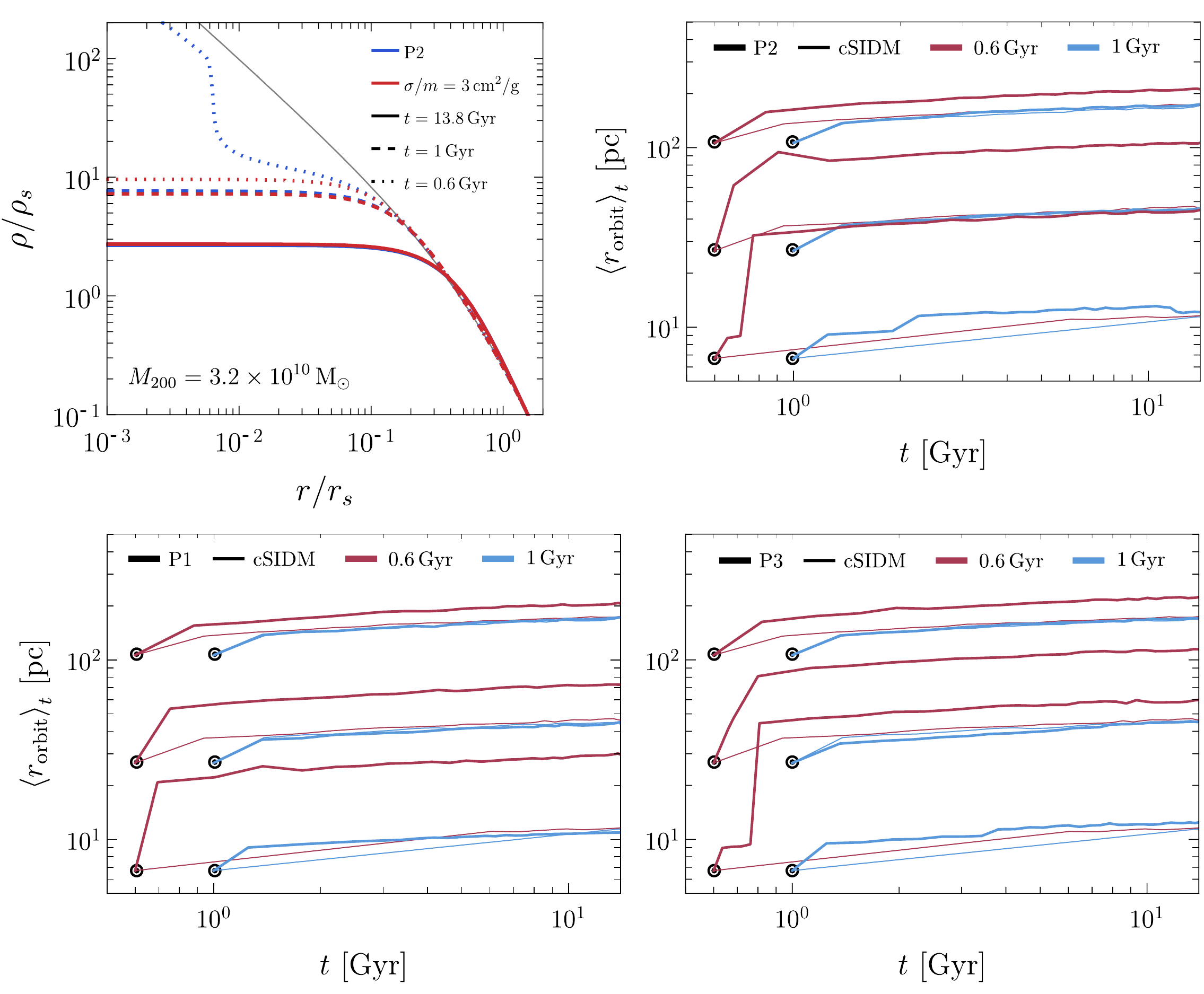}
\caption{
{\bf (Top)} {\it (Left)} - Time evolution of an rSIDM halo ($3.2\times 10^{10}\,{\rm M}_\odot$) in the {\rm P2} benchmark (blue). The time evolution of a cSIDM halo is shown for comparison (red); the cSIDM cross section is chosen so that the cSIDM and the rSIDM benchmark exhibit the same central density at present.
{\it (Right)} - Time evolution of mean orbital radius of stars residing in the rSIDM (thick solid)/cSIDM (thin solid) halos presented in the top-left panel.
The formation time and initial radius of stars are marked by filled circles. For a given initial orbital radius, the distinction between the present orbital radius of stars that have formed just before ($t=0.6\,{\rm Gyr}$, red) and after ($t=1\,{\rm Gyr}$, blue) the thermalization emerges for the rSIDM benchmark, but not for the cSIDM halo.
{\bf (Bottom)} {\it (Left)} - Same as the top-right panel but for the ${\rm P1}$ benchmark. {\it (Right)} - Same as the top-right panel but for the ${\rm P3}$ benchmark.
}
\label{fig:stellarexpand}
\end{figure*}

The bottom-left(right) panel of Fig.~\ref{fig:probes} shows the predicted LOSVD profile with the stellar kinematic parameters for Draco (Segue 1).
The black data points are the observed LOSVD profile.
The NFW profile (gray curve) provides a good fit to the data while the rSIDM benchmarks (blue curves) do not;
this is not surprising since we took best-fit kinematic parameters for the NFW profile.
Note that the presented four curves converge at large radii, i.e., $r\sim r_s$, since the density profiles are identical.
For the Draco-like parameters, one can see that $\sigma_0/m$ controls the central value of $\sigma_{\rm l.o.s.}$;
in the limit of small $\sigma_0/m$, the LOSVD profile in rSIDM will converge to that of the NFW inside the density break $R\lesssim 0.1\,{\rm kpc}$.

Notice that, for the Draco-like parameters, the ${\rm P3}$ benchmark (dotted) exhibits LOSVD curve nearly parallel to the NFW curve for $R\lesssim 0.5\,{\rm kpc}$.
This may motivate one to demonstrate that rSIDM benchmarks can be consistent with the cuspy MW satellites by looking for new stellar kinematic parameter set that makes, e.g., the LOSVD curve in the ${\rm P3}$ benchmark identical to the NFW curve (gray) in the bottom-left panel of Fig.~\ref{fig:probes}.
However, we find that it is hard to match the NFW curve by varying a single stellar kinematic parameter from the best-fit parameter set assuming the NFW profile.
Increasing the assumed mass/concentration of the halo (and thus increase the overall velocity scale) may be another way to uplift the presented LOSVD curve for ${\rm P3}$ to match the NFW curve.
We find that this is not a viable option since increasing the halo mass/concentration puts the velocity scale closer to the resonant velocity, and thus the resultant LOSVD curve is actually lowered due to enhanced core formation in the region $r_{\rm 1,med} \lesssim r \lesssim r_{\rm 1,out}$.
Nevertheless, we remark that rSIDM profiles better fit the observed LOSVD data than cSIDM profiles due to the enhanced central density (see the top-left panel of Fig.~\ref{fig:probes}).
A full parameter scan for rSIDM is beyond the scope of this work.

For halos of mass larger than the specific mass range, the density profile breaks are thermalized.
The thermalization leaves behind a uniform-density core that is virtually indistinguishable from that in cSIDM.
While the explicit smoking-gun signature for rSIDM is absent for these halos, we point out that the orbital radii of stellar tracers that have formed around the thermalization period could serve as indirect evidence for rSIDM.
During the thermalization, there is a rapid outflow of DM near the density break that lowers the central mass density.
The sudden shallowing of the gravitational potential during the thermalization unbinds the stars and increases their orbital radii.
This renders two distinct classes of orbits among the stars that have a similar age.
Stars that formed just before the thermalization experienced a rapid increase in orbital radii upon the propagation of the density break.
Stars that formed just after the propagation experienced the increase that is more marginal and gradual.
The existence of two different classes of stellar orbits is a feature that is distinct between rSIDM and cSIDM since the mass outflow in cSIDM is much more gradual.

To demonstrate the emergence of the two orbital classes from the thermalization dynamics of rSIDM halos, we follow the evolution of mean orbital radius $\langle r_{\rm orbit} \rangle$ of stars that have formed just before and after the passage of the density break.
We focus on a halo mass of $\sim 3.2\times10^{10}\,{\rm M}_\odot$ with median concentration.
For such a halo, the thermalization completes around $t\sim 1\,{\rm Gyr}$~(see the left panel of Fig.~\ref{fig:range}).
We follow the evolution of $\langle r_{\rm orbit} \rangle$ of stars that have formed at $t=0.6\,{\rm Gyr}$ and $1\,{\rm Gyr}$.
We assume the stars to be initially in a circular orbit with an identical orbital radius;
this defines the initial energy and angular momentum of stars.
At a given time $t$, we define the mean orbital radius by a probability weight proportional to the time spent on a line element along the orbit~\cite{Pontzen:2011ty}:
\begin{equation}
\langle r_{\rm orbit} \rangle_t = \int \frac{r \, dr}{\sqrt{E(t)-V_{\rm eff}(r,t)}} \bigg/ \int \frac{dr}{\sqrt{E(t)-V_{\rm eff}(r,t)}} \,,
\label{eq:rorbit}
\end{equation}
where $E$ is the energy of the orbit and $V_{\rm eff}=\Phi+j^2/r^2$ is the 1D effective potential with the specific angular momentum $j$ and gravitational potential $\Phi$.
The angular momentum is conserved throughout the evolution since we consider spherically symmetric halo profiles.
The integration range over $r$ is determined by requiring the integrand to be real.
The time integration is done by assuming the change occurring on timescales shorter than the orbital period of stars is impulsive;
we assume that the change in energy of stars during such timescales is equivalent to the change in $\Phi$ due to evolution of the host halo.
The orbital period is estimated as $\Delta t \sim 2\pi / \sqrt{G \bar{\rho}} \simeq 29\,{\rm Myr}\,\sqrt{(10\,{\rm M_\odot/pc^3})/\bar{\rho}}$ where $\bar{\rho}(t,r)$ is the mean DM density within the radius $r$ at time $t$.
For each time interval, we estimate the mean change in energy of stars as $\langle \Delta E \rangle_t = \langle \Delta V_{\rm eff}\rangle_t$ where the mean over stars is taken as in Eq.~\eqref{eq:rorbit};
when taking the average, we use $E$ and $V_{\rm eff}$ defined at time $t$ while $\Delta V_{\rm eff}=V_{\rm eff}(t+\Delta t)-V_{\rm eff}(t)$.
We update the energy by $\langle \Delta E \rangle_t$ after every time step.

The evolution of $\langle r_{\rm orbit} \rangle_t$ for each rSIDM benchmark is shown in the top-right and the bottom panels of Fig.~\ref{fig:stellarexpand};
we display the evolution of stars that formed just before (red) and after (blue) the thermalization.
For each benchmark, the distinction in the present $\langle r_{\rm orbit} \rangle_t$ between the two is highlighted towards a smaller initial radius, due to a more significant change in enclosed mass upon the passage of the density break.
For a given initial radius, the distinction is larger for smaller $\sigma_0/m$ since the central DM density before the thermalization is larger for smaller $\sigma_0/m$ while the density profiles after the thermalization are the same for all the benchmarks.
Such distinctions are not present for cSIDM halos.
We also display the evolution $\langle r_{\rm orbit} \rangle_t$ for cSIDM halos as thin curves in the top-right and the bottom panels of Fig.~\ref{fig:stellarexpand}.
We choose a constant SIDM cross section $\simeq 3\,{\rm cm^2/g}$ so that the central density at present is the same as that of the rSIDM benchmarks; see the top-left panel of Fig.~\ref{fig:stellarexpand}.
Although the cSIDM has the same present density profile as rSIDM, the two distinct classes of stellar orbits only emerge for the rSIDM benchmarks due to the intrinsically different evolution history of rSIDM halos.

\section{Conclusions} \label{section:conclusion}

It has been recently demonstrated that the inferred structures of ultra-faint dwarf galaxies can put a stringent upper bound on the SIDM cross section.
If we take the constraint at faces values, any SIDM model invoked to address the core-cusp problem of rotation-supported galaxies should have an extraordinary velocity dependence where $\sigma/m$ sharply drops by an order of magnitude towards lower velocities around the scattering velocity of $\sim 30\,{\rm km/s}$.
Resonant SIDM (rSIDM) is an interesting possibility that exhibit such velocity dependence.
To further assess this possibility, structural evolution of resonant SIDM halos needs to be explored and compared to that in constant-cross section SIDM (cSIDM).

In this work, we have studied the structural evolution of isolated resonant SIDM halos for the first time by employing the gravothermal fluid method.
We have focused on the three resonant SIDM benchmarks of different offset cross sections that may explain the low central densities of dwarf/LSB galaxies while being consistent with the constraints.
We remark that while the benchmarks with smaller values of the offset cross section, e.g., the ${\rm P2}$ and ${\rm P3}$ benchmarks, better evades the constraints from the UFD's, the high-velocity limit of the cross section hardly fit the cross section favored by central cores of galaxy clusters.
For each benchmark, we have scoped the halo-mass dependence of the structural evolution over a wide mass range, i.e., $10^9$-$10^{15}\,{\rm M_\odot}$.

We have found that, except for the halos in a specific mass range, present-day structures of rSIDM halos are virtually indistinguishable from that in cSIDM.
For the halos smaller than the specific mass range, the DM scattering velocities do not overlap with the resonant velocity range at any radius and thus their structure is determined by the constant offset cross section.
For the halos within the specific mass range, only a limited region inside a halo experiences the resonance and develops a density break that is observable in their present structures.
For the halos larger than the range, the density break is thermalized before the present and their structure becomes indistinguishable from that of cSIDM halos.
We have delineated the halo-mass range for each rSIDM benchmark from the simulations.
We have found that the low-end of the mass range can be conveniently predicted by analyzing the radial distribution of the heat conduction timescale of the initial NFW halo, while the high-end of the range can be over-estimated by this method.

The density break of the rSIDM halos in the specific mass range may be explicitly probed by astrophysical observations.
We have demonstrated how the density break can be manifested in rotation curves and LOSVD profiles.
The density break renders a break in the rotation curve that is more highlighted for smaller values of the offset cross section.
We have also demonstrated how the rSIDM benchmarks can better fit the LOSVD profiles of some of the most cuspy satellites of the MW compared to cSIDM, while explaining the low central densities of rotation supported galaxies. This motivates us to do a concrete investigation on how the observed structures of galaxies constrain the rSIDM parameter space, while we leave it for future work.
Lastly, we have suggested a distinct observable feature of rSIDM halos that are thermalized at present.
The thermalization dynamics of the density break renders two distinct orbital classes among stars with a similar age, depending on whether a star has formed before or after the thermalization;
the former would have significantly larger orbital radius than the latter at present.
We have demonstrated that such a feature can be used to distinguish rSIDM from cSIDM in principle.

Although we have demonstrated interesting dynamics of isolated rSIDM halos, confirming if similar dynamics and observable signatures are valid for cosmological halos is warranted.
Once confirmed, having a semi-analytic modeling for rSIDM halo structures would be helpful for searching the smoking-gun signatures of rSIDM in astrophysical observations.
It would also be interesting to check if the resonant self-interaction contributes to the diversity of halo structures in cosmological simulations.
For example, we have demonstrated the diverse structures among rSIDM halos of similar mass;
while the halos in a specific mass range still exhibit large central density, halos slightly above the range could exhibit significantly lower central density due to the thermalization of the density break.

\acknowledgments{The authors would like to acknowledge Kohei Hayashi and Satoshi Shirai for providing the line-of-sight velocity dispersion data and the corresponding best-fit structural parameters. A. K. acknowledges partial support from Norwegian Financial Mechanism for years 2014-2021, grant nr 2019/34/H/ST2/00707; and from National Science Centre, Poland, grant 2017/26/E/ST2/00135 and DEC-2018/31/B/ST2/02283. H.K. thanks Hyungjin Kim and Chang Sub Shin for helpful discussions. The work of H.K. is supported by IBS under the project code, IBS-R018-D1.}

\appendix

\section{Velocity-averaged cross section}
\label{appendix:averages}

The distribution average of the rSIDM cross section [see Eq.~\eqref{eq:BWparam}] with respect to the Maxwell-Boltzmann distribution is given as
\begin{equation}
\frac{\left\langle \sigma v_{\rm rel}\right\rangle}{m}=\frac{\sigma_{0}\langle v_{\rm rel}\rangle}{m}+\frac{256\pi S}{m^3}{\cal I}_{L}\left(\gamma,v_{R},\nu \right)\,,
\label{eq:sigamvappendix}
\end{equation}
where the integral ${\cal I}_{L}$ is defined as
\begin{equation}
\begin{aligned}
{\cal I}_{L}\left(\gamma,v_{R},\nu\right)=\int_{0}^{v_{{\rm max}}}&\frac{\gamma^{2}v_{{\rm rel}}^{4L+1}}{\left(v_{{\rm rel}}^{2}-v_{R}^{2}\right)^{2}+16\gamma^{2}v_{{\rm rel}}^{2(2L+1)}}\\
&\times f(v_{\rm rel};\nu) \, dv_{{\rm rel}}\,.
\end{aligned}
\label{eq:IL}
\end{equation}
As in the main text, we take the local escape velocity to be infinity.
At the vicinity of the resonant velocity, i.e., $\langle v_{\rm rel} \rangle = (4/\sqrt{\pi}) \nu \sim v_R$, ${\cal I}_L$ is dominated by the singular behavior of the denominator in Eq.~\eqref{eq:IL}.
One may approximate such a contribution using the narrow-width approximation (NWA);
using $\underset{\epsilon\rightarrow 0}{\lim} \, \frac{\epsilon}{x^2+\epsilon^2}=\pi\delta(x)$, we make the replacement
\begin{equation}
\frac{1}{\left(v_{{\rm rel}}^{2}-v_{R}^{2}\right)^{2}+16\gamma^{2}v_{{\rm rel}}^{2(2L+1)}}\rightarrow \frac{\pi\delta (v_{\rm rel}-v_R)}{8\gamma v_R^{2(L+1)}}\,,
\label{eq:Idenom}
\end{equation}
which leads to the resonant contribution given in Eq.~\eqref{eq:NWA} of the main text.
One can find that the peak value of the resonant contribution is given as $96\sqrt{6}\pi^{3/2}S\gamma v_R^{2L-2}/(e^{3/2} m^3)$ which is only $\gamma$-suppressed due to the cancelation of a $\gamma$ from the NWA.
Note that the NWA becomes more accurate as the half width of the distribution Eq.~\eqref{eq:Idenom} is smaller compared to $v_R$, i.e., $\Delta v_{\rm rel}\sim 2v_R^{2L}\gamma < v_R$.

\begin{figure*}[t]
\centering
\includegraphics[width=1\textwidth]{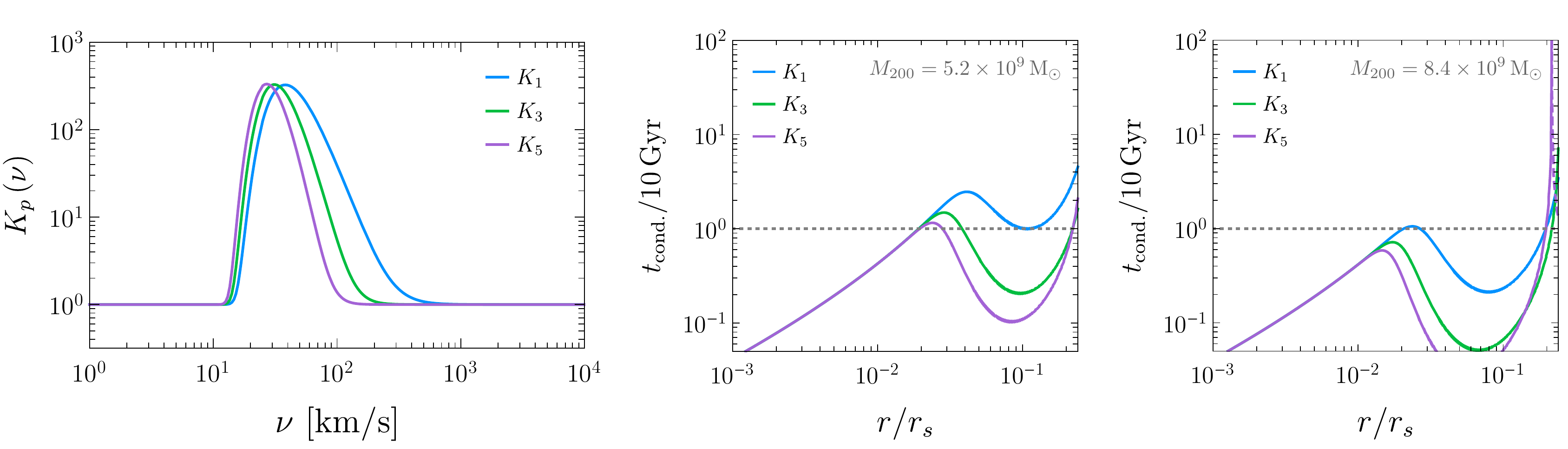}
\caption{
{\bf (Left)} - Velocity dependence of $K_p(\nu)$ for different values of $p$.
For $p=1$, the rSIDM parameters are the same as the ${\rm P2}$ benchmark.
For $p=3$ and $p=5$, the rSIDM parameters are the same as the ${\rm P2}$ benchmark except the DM mass is adjusted so that the maximum of $K_p$'s are identical to that of $p=1$;
$m/S^{1/3}=0.4\,{\rm GeV}$ for $p=1$, $m/S^{1/3}=0.43\,{\rm GeV}$ for $p=3$ and $m/S^{1/3}=0.45\,{\rm GeV}$ for $p=5$.
{\bf (Middle)} - Radial profile of $t_{\rm cond.}$ for different values of $p$. The presented halo mass is the low end of the specific mass range (defined by taking $p=1$ for the distribution averaging of the SIDM cross section) for the ${\rm P2}$ benchmark, i.e., $M_{200}=5.2\times 10^9\,{\rm M}_\odot$.
{\bf (Right)} - Same as the middle panel, but the presented halo mass is the high end of the specific mass range for the ${\rm P2}$ benchmark, i.e., $M_{200}=8.4\times 10^9\,{\rm M}_\odot$.
}
\label{fig:Kp}
\end{figure*}

We define the contributions to ${\cal I}_L$ outside the width as the out-of-pole contributions.
For the contribution from the low-velocity side, i.e., $0\leq v_{\rm rel}\lesssim v_R-2v_R^{2L}\gamma$, one may approximate
\begin{equation}
\frac{1}{\left(v_{{\rm rel}}^{2}-v_{R}^{2}\right)^{2}+16\gamma^{2}v_{{\rm rel}}^{2(2L+1)}}\rightarrow \frac{1}{v_{R}^4}\,.
\end{equation}
With this replacement, the out-of-pole contribution in the low-velocity side is approximated as
\begin{align}
{\cal I}_{L,{\rm low}}\left(\gamma,v_{R},\nu\right)&=\int_{0}^{v_R-2v_R^{2L}\gamma}\frac{\gamma^{2}v_{{\rm rel}}^{4L+1}}{v_R^4}f(v_{\rm rel};\nu)dv_{{\rm rel}} \nonumber \\
&=\frac{4^{2L+1}\sqrt{\pi} \gamma^2}{v_R^4}\nu^{4L+1}\\
&\times\left[\Gamma\left(2L+2\right)-\Gamma\left(2L+2,\frac{v_R-2v_R^{2L}\gamma}{4\nu^2}\right)\right]\,, \nonumber
\end{align}
where in the second equality, the second term in the square bracket is the incomplete Gamma function which is negligible in the $\nu\rightarrow 0$ limit.
Away from the resonant velocities, i.e., $\langle v_{\rm rel} \rangle \ll v_R$, this out-of-pole contribution is dominant over the resonant contribution which is exponentially suppressed.
Remark that in the $s$-wave scattering case, ${\cal I}_{L,{\rm low}}$ is only $\nu$-suppressed which is the same suppression for the off-set contribution (the first term) in the RHS of Eq.~\eqref{eq:sigamvappendix};
one needs to ensure that this constant out-of-pole contribution to $\langle\sigma v_{\rm rel}\rangle/(m\langle v_{\rm rel}\rangle)$ is small enough to be compatible with the constraints from the UFDs.

Similarly for the high-velocity side, i.e., $v_{\rm rel}\gtrsim v_R+2v_R^{2L}\gamma$, we make the replacement 
\begin{equation}
\frac{1}{\left(v_{{\rm rel}}^{2}-v_{R}^{2}\right)^{2}+16\gamma^{2}v_{{\rm rel}}^{2(2L+1)}}\rightarrow \frac{1}{v_{\rm rel}^4}\,,
\end{equation}
which approximates the out-of-pole contribution as
\begin{equation}
\begin{aligned}
{\cal I}_{L,{\rm high}}\left(\gamma,v_{R},\nu\right)=&\,4^{2L-1}\sqrt{\pi}\gamma^2\nu^{4L-3}\\
&\times\Gamma\left(2L,\frac{(v_R+2v_R^{2L}\gamma)^2}{4\nu^2}\right)\,,
\end{aligned}
\end{equation}
where the incomplete Gamma function converges to the complete Gamma function in the high-$\nu$ limit.
Inspecting the $\nu$-dependence of ${\cal I}_{L,{\rm high}}$, this out-of-pole contribution dominates over the resonant contribution for $p$-wave scattering in the high-$\nu$ limit.
For the $s$-wave scattering, the $\nu$-dependencies are the same between the resonant and the out-of-pole contributions, and the relative magnitude between them depends on the assumed rSIDM parameters.

\section{Distribution averaging in rSIDM}
\label{appendix:Kp}

In the main text, we have assumed a specific distribution averaging for the SIDM cross section appearing in the thermal conductivity in the LMFP regime, i.e., $\kappa_{\rm LMFP}\propto K_1(\nu)$.
Since the halo evolution we have explored resides in the LMFP regime, our quantitative results, e.g., the specific mass range of rSIDM halos, may change among different choices for the distribution averaging.
The velocity dependence of $K_p(\nu)$ for different choices of $p$ are shown in the left panel of Fig.~\ref{fig:Kp};
we took the ${\rm P2}$ benchmark for $p=1$, and adjusted the DM mass from the ${\rm P2}$ benchmark for $p=2,\,3$ so that the peak values of $K_p$ are identical.
The velocity range where the $K_p$ is resonantly enhanced is narrower for larger values of $p$ since $\langle v^p_{\rm rel} \rangle \propto \nu^p$, the denominator of $K_p$, exhibits sharper velocity dependence for $\langle v_{\rm rel} \rangle > v_R$ while the numerator barely depends on $\nu$ (predominantly depend on $v_R$).

The difference in the velocity dependence may change the specific mass range for rSIDM halos.
The expected change is a slight shift in the low/high end towards a smaller halo mass.
In the middle panel of Fig.~\ref{fig:Kp}, we show the $t_{\rm cond.}$ profiles inside a halo of mass $M_{200}=5.2\times 10^9\,{\rm M}_\odot$, which is the low-end of the specific mass range presented in the main text, i.e., for $p=1$ (determined by the condition $t_{\rm break}=10\,{\rm Gyr}$).
For larger values of $p$, $t_{\rm cond.}$ is relatively shorter around $r=r_{\rm break}$ due to the sharper velocity dependence of $K_p$.
Therefore, the estimation for the low end of the specific mass range is smaller for larger values of $p$;
the low end determined by $t_{\rm break}=10\,{\rm Gyr}$ is $4 \,(3)\times 10^9\,{\rm M}_\odot$ for $p=3\,(5)$.
Similarly, the high end of the specific mass range may be smaller for larger values of $p$; see the right panel of Fig.~\ref{fig:Kp}.
The high-end mass estimated by the condition $t_{\rm therm}=10\,{\rm Gyr}$ is $6 \,(5)\times 10^9\,{\rm M}_\odot$ for $p=3\,(5)$, while $8.4\times 10^9\,{\rm M}_\odot$ for $p=1$.

\section{Cored NFW initial conditions}

\begin{figure*}[t!]
\centering
\includegraphics[width=0.93\textwidth]{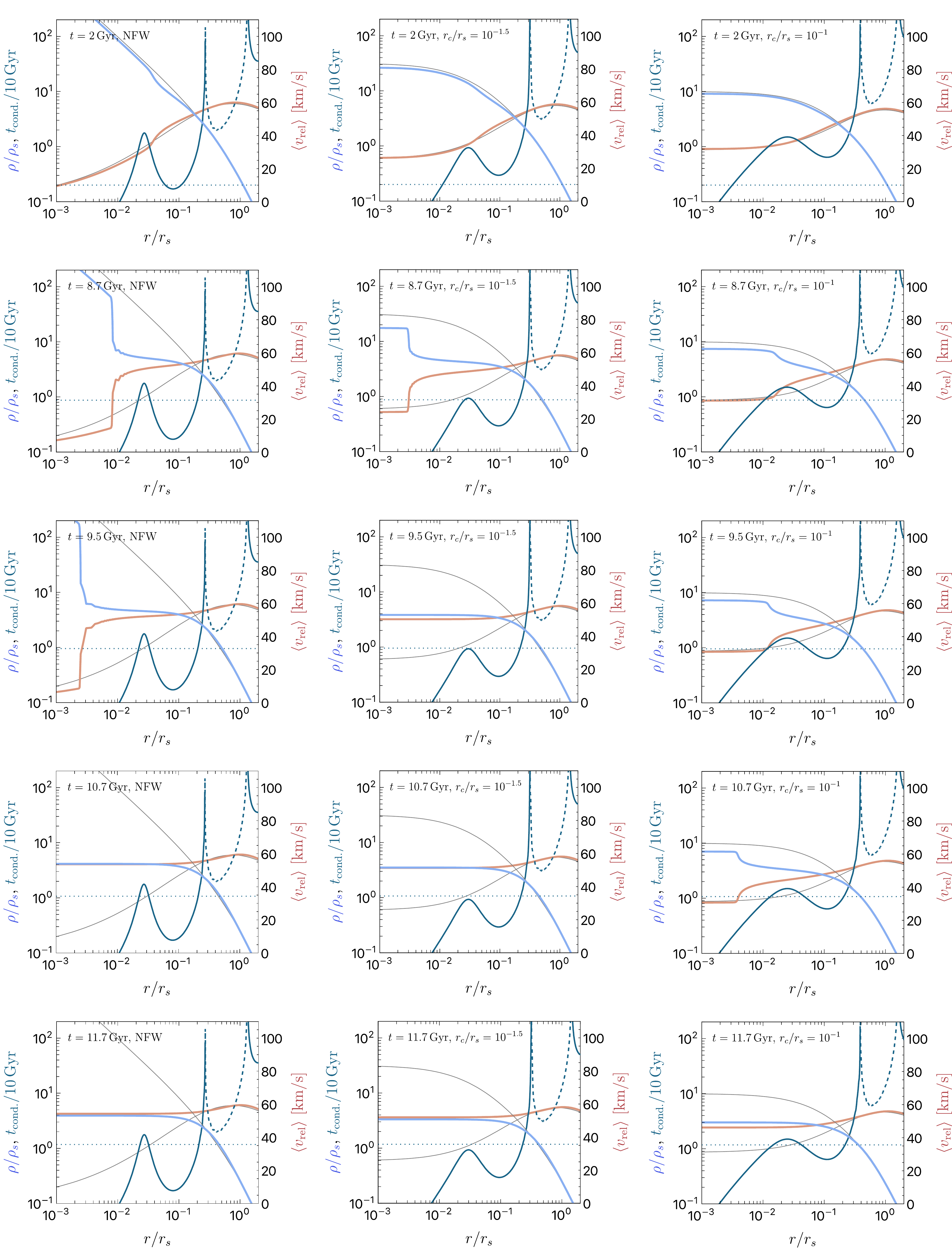}
\caption{
Snapshots for the time evolution of a rSIDM halo of mass $M_{200}=10^{10}\,{\rm M}_\odot$ in the ${\rm P3}$ benchmark;
see the caption of Fig.~\ref{fig:break} in the main text for the information of each curves.
The left panels assume the NFW profile (gray curves) as the initial condition.
The middle (right) panels assume the cored NFW profile (gray curves) with $r_c/r_s=10^{-1.5}$ ($r_c/r_s=10^{-1.}$) as the initial condition.
}
\label{fig:CNFW}
\end{figure*}

In the main text, we have assumed the NFW profile as the initial condition for rSIDM halos.
However, finite initial core densities may be rendered in reality due to, e.g., the astrophysical processes of baryons.
In this section, we try to scope the possible impact of initial cores on the distinctive evolution of rSIDM halos, i.e., the formation and thermalization of the density break.
We take the cored NFW profile (CNFW) to represent the initial cores of rSIDM halos:
$\rho_{\rm cnfw}(r)=\rho_s\,r_s/[(r+r_c)\left(1+r/r_s\right)^2]$ where $r_c$ parameterizes the initial core size.
In Fig.~\ref{fig:CNFW}, we present the time evolution of rSIDM halos with different values of the initial core size in the ${\rm P3}$ benchmark;
the left panels are for the NFW profile ($r_c=0$), and the middle (right) panels are for the finite initial core size $r_c/r_s=10^{-1.5}$ ($r_c/r_s=10^{-1}$).

The formation time of the density break can be conveniently estimated by $t_{\rm break}$ which is the time at which the halo age is identical to the local minimum of the $t_{\rm cond}$ profile, as discussed in Section~\ref{section:rSIDMevolution} of the main text.
The values of $t_{\rm break}$ are $1.7\,{\rm Gyr}$, $2.9\,{\rm Gyr}$ and $6.2\,{\rm Gyr}$ for left, middle and right panels, respectively;
the formation time of the density break is smaller for smaller initial core size.
This is because initial profiles with larger cores have smoother central profiles which render longer heat conduction timescale.
Indeed, the density break appears later for halos with larger initial cores, as demonstrated in the snapshots for $t=2\,{\rm Gyr}$ in Fig.~\ref{fig:CNFW}.

Contrary to the formation time of the density break, the thermalization time of rSIDM halos does not display a clear trend with respect to the initial core size.
The values of the thermalization timescale $t_{\rm therm.}$ estimated from the initial $t_{\rm cond}$ profile are $18\,{\rm Gyr}$, $9.4\,{\rm Gyr}$ and $15\,{\rm Gyr}$ for left, middle and right panels, respectively.
The halo with $r_c/r_s=10^{-1.5}$ (middle panels) has a shorter thermalization timescale than the halo of the initial NFW profile (left panels) since the SIDM cross section is resonantly enhanced due to larger scattering velocities at the center.
Such tendency is reversed as we further increase the core size;
the halo with $r_c/r_s=10^{-1.}$ (right panels) has a longer thermalization timescale than the halo with $r_c/r_s=10^{-1.5}$ (middle panels).
This is because when the initial core size becomes comparable to $r_{\rm break}$, i.e., the radial position of the local minimum of the initial $t_{\rm cond}$ profile, the central profiles become smooth enough and render longer heat conduction timescale for larger initial core size.
Meanwhile, the actual thermalization time of the density breaks is not identical to our estimated timescales $t_{\rm therm.}$, as we have discussed in Section~\ref{section:rSIDMevolution} of the main text.
The actual thermalization times are $9.5\,{\rm Gyr}$, $8.7\,{\rm Gyr}$ and $10.7\,{\rm Gyr}$ for left, middle and right panels, respectively.
The trend based on the comparison among $t_{\rm therm.}$ coincides with the comparison of the actual thermalization times between the left and middle panels, and between the middle and right panels, with an exception for the comparison between the left and right panels.
This suggests that comparison among $t_{\rm therm.}$ is not enough to capture the trends of the variation of the thermalization time with respect to the initial core size.
Nevertheless, our results shows that the thermalization time has weak dependence on the initial core size.

\clearpage
\bibliographystyle{utphys}
\bibliography{rSIDM}

\end{document}